\documentclass[aip, pop,  showpacs,amsmath,amssymb]{revtex4-1}
\usepackage{amssymb}

\usepackage{txfonts}

\usepackage[pdftex,colorlinks=true,bookmarks=false,citecolor=blue,urlcolor=blue]{hyperref} %latex w/dvips

\usepackage{graphicx}% Include figure files
\usepackage{dcolumn}% Align table columns on decimal point
\usepackage{bm}% bold math

\newcommand{\new}[1]{{ \color{red} #1}}

\usepackage{CJKutf8}

\usepackage{tikz}

\begin{document}

\begin{CJK}{UTF8}{gbsn}

\title{Space-charge effect of the time-varying electron injection in a diode: classical and relativistic regimes\footnote{Drafted November 2016, revised 18 April 2017, POP51880 submitted 20 April, review received 1 May, revised 28 June and 16 July, POP51880A resubmitted 16 July, minor rev received 26 July, POP51880B re-resubmitted 29 July, accepted 3 Aug 2017. }}

\author{Yangji\'e Liu (刘泱杰)}
\email[Electronic mail of corresponding author: ]{yangjie@hubu.edu.cn}
%\affiliation{School of Electronics and Information, Hangzhou Dianzi University, Xiasha, Hangzhou, P. R. China}
\affiliation{Faculty of Physics and Electronic Science, Hubei University, and Hubei Collaboration Innovation Centre for Advanced Organic Chemical Materials -- Hubei Key Laboratory of Ferro \& Piezoelectric Materials and Devices, 430062 Wuhan, Hubei Province, P. R. China}
\affiliation{Antennas Group, School of Electrical Engineering and Computer Science, Queen Mary University of London, E1 4NS Mile End, East London, United Kingdom}

\author{Qi Tang}
%\email{tangq3@rpi.edu}
\affiliation{Department of Mathematical Sciences, Rensselaer Polytechnic Institute, United States}

\author{Biyi Wu (吴比翼)}

\affiliation{Antennas Group, School of Electrical Engineering and Computer Science, Queen Mary University of London, E1 4NS Mile End, East London, United Kingdom}
\affiliation{Center for Electromagnetic Simulation, Beijing Institute of Technology, Haidian District, Beijing, P. R. China}

%\author{Wenfeng Wang (王文峰)}
%\email{w.wang@hubu.edu.cn}
%\affiliation{Faculty of Physics and Electronic Science, Hubei University, 430062 Wuhan, P. R. China}

\date{compiled \today, submitted 29 July, accepted 3 Aug 2017 to \emph{AIP Phys. Plasmas}}
%submitting to arXiv and \emph{AIP Phys. Plasmas}.

%\thanks{Author to whom correspondence should be addressed}

\begin{abstract}
\noindent 

% Our calculation shows that it is \emph{unlikely} that one can emit a time-varying injection current from cathode to transmit more than the known space-charge limits, in either classical and relativistic regime. 

Conventionally, space-charge (SC) limited current density is defined as the maximal current density allowed to traverse a diode under a DC voltage when a time-invariant flow is injected from the cathode. In this work, we study the SC limited current density under a time-varying injection for both classical and relativistic regimes and determine the maximal amount of limited current density under certain conditions. Our simulations show that it is unlikely that a time-varying injection emitted from cathode exceeds the known SC limits, in either classical or relativistic regime.

\end{abstract}

\pacs{52.59.Sa, 71.15.-m, 52.65.Rr, 68.43.Tj}

% Particle-in-cell method (plasma simulation), 52.65.Rr

% Desorption field induced, 79.70.+q

% Photon stimulated desorption (see also 79.20.La Photon- and electron-stimulated desorption)
% 68.43.Tj

% Particle beams    
%intense beams in plasma, 52.59.-f

%Space-charge-dominated beams (plasmas), 52.59.Sa

%condensed matter
%calculation methods, 71.15.-m

%Nanoparticles
%electronic structure of, 73.22.-f

%Electron emission
%thermionic emission, 79.40.+z

%suggesting reviewers: Koh, Zhang P, Zhu Y and Lin(Tex-Vorpal)

\maketitle

\end{CJK}

\section{Introduction}
It is of practical interest for plasma devices to seek enhanced density of plasma sources, which typically requires consideration of space-charge (SC) limit for a diode structure. SC limited current density describes the maximal current density allowed to transport through a diode, for a time-invariant injection from the cathode. For a diode with a gap spacing of $d$ and a diode voltage of $\phi_g$, this is known as the well-known Child-Langmuir law~\cite{Child1911, Langmuir1913, ZhangP2017}. This law  extends to the more general case of nonzero initial velocity $v_0$~\cite{Jaffe1944, LiuS1995} and the relativistic regime~\cite{Jory1969, Howes1966b, Feng2008, Li2009, YL2014} when the voltage potential becomes relativistic. In recent years the efforts to achieve a higher electron flow have been very active, in both classical and relativistic regimes~\cite{Luginsland2002, Kumar2008}. Note that self-magnetic effect and inductive effect can both be additional factors to suppress SC limit~\cite{Kumar2008}.

There has been recent interest in investigating whether an injection of time-varying current density can contribute to a larger current density that is transmittable across the diode space~\cite{Caflisch2012, Griswold2010, Zhu2011, Griswold2012, Liu2012b, Pant2013, Liu2014, MyThesis, Griswold2016}. On the other hand, time-varying current emission through the diode anode arises naturally even when the injection current density is \emph{time-invariant}, with the diode voltage being fixed. For example, a recent work based on 1D charge sheet model investigated how the temporal interval becomes distorted due to space-charge effect~\cite{YL2015}. Another recent work~\cite{Caflisch2012} used a characteristic method to solve the 1D pressureless Euler-Poisson equations and achieved an over-classical limit by varying the anode boundary voltage with time, although the relevant characteristics are not solvable in a closed form for arbitrary temporal profiles of injection. Therefore, a general treatment for time-varying flow of electrons is demanded. In this work, we address the practical issue of time-varying injection current by solving the same 1D pressureless Euler-Poisson equations, in order to unfold the physics of time-varying injection flow of electrons. According to our simulation, it seems \emph{unlikely} that one can transmit a time-varying injection $j(t)$ across the diode that is time-averagely higher than the conventional SC limited current density, in both classical~\cite{Jaffe1944} and relativistic regimes~\cite{Feng2008, ZhangY2009}. 
In our paper, the electrons are injected into the diode gap following the prescribed time profiles. It is assumed that SC limit is reached when reflection occurs~\cite{Chen2011}. Then the average current densities flowing through the anode are computed and compared to the previous known SC limits. This approach is also extended to the relativistic regime~\cite{Feng2008} when the diode voltage or the initial velocity become relativistic, and the same negative result prevails.

\section{Problem and method}\label{Prob}

In this section, we describe the problem of electron flow in a 1D diode structure depicted in Fig.~\ref{fig:Fig1} in the classical regime, and the numerical approach to determine the SC limit. They are also extended to the relativistic regime when the initial velocity becomes comparable to the speed of light, or when the relativistic diode voltage becomes comparable with the rest energy of electrons. 

\subsection{Classical regime}\label{Classical}
Consider a diode with a gap of spacing $d$ and a fixed boundary potential difference $\phi_g$ between the cathode and anode. The electrons are injected from the cathode ($x=0$) with a forced time-varying current density of $j_0(t)$\footnote{In 1D case, the current density $j$ are simply related to the current $J$: $j=J/A$ where $A$ is the area. } over a pulse length $\tau_{p}$. We take the pulse length $\tau_{p}$ to be longer than the transmit time. 
The 1D pressureless Euler-Poisson equations for the electron flow in a diode can be written as
\begin{align}
\partial_t\rho+\partial_x(\rho v) & =0, \\
\partial_t v+v\partial_x v & =\partial_x\phi,\\
\partial_x^2\phi & =\rho, 
\end{align}
in which scaled quantities are used for position $x$, time $t$, velocity $v$, potential $\phi$ and charge density $\rho$ similar to the previous work~\cite{Caflisch2012}. Note that all primed ones variables in SI units are scaled into unprimed ones to simplify the equations:
\begin{align}
\label{scaling}
(x, t, v, \phi, \rho)=\Big(\frac{x'}{L}, \frac{t'}{L}, \frac{v'}{L/T}, \frac{\phi'}{\Phi}, \frac{\rho'}{R}\Big), 
\end{align}
where $L, T$ are the characteristic length and time,  and the characteristic potential and density are $\Phi=mL^2/(eT^2), R=\epsilon_0\Phi/eL^2$ for simplicity~\cite{Caflisch2012}. An inflow boundary condition is used at $x=0$,
\begin{align}\label{inflow}
    \phi=0,  \quad v = v_0,\quad  \rho(t) =\frac{J_0(t)}{v_0}, \qquad 0\le t \le \tau_p,
\end{align}
to take into account the injection current density $J_0(t)$ at the cathode $x=0$ (see Fig.~\ref{fig:Fig1}).  Note that for a special case when $v_0=0$, our algorithm can deal with it with a different inflow boundary condition. More details will be elaborated in Subsec.~\ref{nullv}. The outflow boundary condition used at $x=d$ is a fixed potential $\phi = \phi_g$.

{\newcommand{\lbfont}{\small}
\def\xL{.75}
\def\ysb{5}
\def\ysa{0}

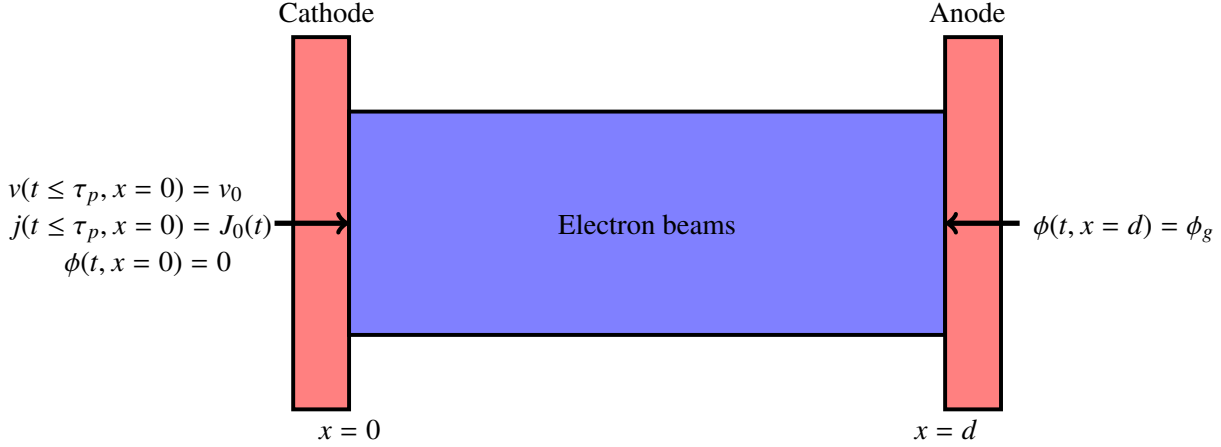
\begin{figure}[hbt]
\begin{center}
 \resizebox{10cm}{!}{% START resize box
 \begin{tikzpicture}[scale=1]
  \useasboundingbox (0,0) rectangle (10,5);  % set the bounding box (so we have less surrounding white space)
  %
  % ---- translating disk ----
  \begin{scope}[xshift=0cm,yshift=-0.5cm]
   \begin{scope}[xshift=.25cm]
      \draw[thick,fill=red!50,draw=black,line width=1.5pt] (0,0) rectangle (\xL,\ysb);
    \end{scope}
   \begin{scope}[xshift=1cm,yshift=1cm]
      \draw[thick,fill=blue!50,draw=black, line width=1.5pt] (0,0) rectangle (8,\ysb-2);
    \end{scope}
    \begin{scope}[xshift=9cm]
      \draw[thick,fill=red!50,draw=black, line width=1.5pt] (0,0) rectangle (\xL,\ysb);
    \end{scope}
    \draw (1  ,0) node[anchor=north,black,yshift=0pt] {\lbfont$x=0$};
    \draw (9  ,0) node[anchor=north,black,yshift=0pt] {\lbfont$x=d$};
    \draw[->,line width=2pt,xshift=0pt] (0,2.5) -- (1,2.5) ;
    \draw[<-,line width=2pt,xshift=0pt] (9,2.5) -- (10,2.5) ;
    %the boundary condition at x=d
    \draw (11.4 ,2.75) node[anchor=north,black,yshift=0pt] {\lbfont $\phi(t,x=d) = \phi_g$};
    %the boundary condition at x=0
    \begin{scope}[yshift=-0.5cm]
    \draw (-2 ,3.75) node[anchor=north,black,yshift=0pt] {\lbfont $v(t\le \tau_p,x=0)=v_0$};
    \draw (-1.8 ,3.25) node[anchor=north,black,yshift=0pt] {\lbfont $j(t\le \tau_p,x=0)=J_0(t)$};
    \draw (-1.7 ,2.75) node[anchor=north,black,yshift=0pt] {\lbfont $\phi(t,x=0) = 0$};
    \end{scope}
%    %the arrows on the bottom (I do not like it)
%    \draw[<->,very thick] (1,0.2) -- (9,0.2);
%    \draw (5  ,.75) node[anchor=north,black,yshift=0pt] {\lbfont$d$};
    \draw (5  ,2.75) node[anchor=north,black,yshift=0pt] {\lbfont Electron beams};
    \draw (.7  ,5.6) node[anchor=north,black,yshift=0pt] {\lbfont Cathode};
    \draw (9.3 ,5.6) node[anchor=north,black,yshift=0pt] {\lbfont Anode};
  \end{scope}
  %we can turn on grid for measurement (it will make changing those position numbers much easier)
  %\draw[step=1cm,gray] (0,0) grid (10,5);
  %
  \end{tikzpicture}
}% end of resize box

\end{center}

\caption { \label{fig:Fig1} Schematic: 1D diode model of voltage difference $\phi_{\rm g}$ and spacing $d$. Time-dependent injection current density $J_0(t)$ up to finite time $0\leq t\leq \tau_{\rm p}$ emits from the cathode, and is dragged towards the anode with initial velocity $v_0$. The blue color indicates the electron beams in the diode spacing.  }

\end{figure}
}

The space-charge limited current density for time-invariant injection in scaled quantity can be written as~\cite{Jaffe1944, Caflisch2012}
\begin{equation}
\label{Jm}
J_{\rm m}=\frac{2}{9d^2}\Big(v_0+\sqrt{v_0^2+2\phi_g}\Big)^3. 
\end{equation}
For arbitrary $J_0(t)$ at cathode, a Lax-Friedrichs scheme coupled with a second-order Poisson solver provides insight into solving the system numerically in \texttt{MATLAB}. In Sec.~\ref{Results} the numerical solutions are used to observe how the charge density and diode potential along diode space $d$ vary with time. The magnitude of the injection density is increased from small to large, and we determine that SC limit is reached when negative velocity \emph{just} appears near cathode under injection of a certain emission profile. Here we assume that emission profile $J_0(t)$ can be expressed as 
\begin{equation}
J_0(t)=\beta_mJ_m\cdot j_m(t), 
\end{equation}
where $\beta_m$ is the dimensionless magnitude and $j_m$ is a particular time-dependent profile assumed as follows, 
\begin{align}
j_0& =1, \\
j_1& =2\bar{t},\\
j_2& =3\bar{t}^2, \\
j_3& =2\big(1-\bar{t}\big),\\
j_4& =3-3\bar{t}^2,\\
j_5& =-6\bar{t}^2+6\bar{t},\\
j_6& =12\left(\bar{t}-\frac{1}{2}\right)^2,
\end{align}
where the time is normalised by $\bar t=t/\tau_{\rm p}$. Compared with other temporal profiles, these polynomial emission profiles are chosen based on the understanding that a smooth temporal function should be expanded as a series of polynomials according to Taylor series. We first choose a steady function (8), then monotonic increasing functions (9, 10), monotonic decreasing functions~(11, 12), and lastly a valley function (13) and a mountain function~(14)~\footnote{The increasing profiles (8-10) are normalised so that within the emission pulses they give the same account of charge $\int_0^{\tau_{\rm p}}j_m(t){\rm d}t=\int_0^{\tau_{\rm p}}1{\rm d}t=\tau_{\rm p}$ and the other profiles are chosen to be as smoothly-varying as possible. }. We sweep magnitudes $\beta_m$ for each $m$ in profile $j_m(t)$ and decide the SC limit is reached when reflection \emph{just} occurs, i.e. the velocity becomes negative, within the pulse length $0\leq t\leq  \tau_{\rm p}$. As a reference when the injection flow is time-invariant $m=0$, it is understood that $\beta_0$ should be one which corresponds to the convectional SC limit $J_m$ of Eq.~\eqref{Jm}, while other $\beta_m$'s generally differ from one. 

In our simulations, we choose the electron beam length to be longer than the typical transmit time of an electron across the spacing~(cf. Figs. 2-3 below). This is for the following reason. In empirical situations, one can only inject an electron beam of finite length. We thus do so in order to understand the sole role of time-dependent injection. This long pulse condition is important as when the temporal length of the electron beam becomes shorter than Child-Langmuir transmit time $3d/\sqrt{2\phi_g}$,  the short-pulse limited current density can transcend Child-Langmuir limit already~\cite{Valfells2002}. Therefore we intensionally implement this long pulse condition in order to remove complication resultant from the short-pulse effect~\cite{Valfells2002}.

\subsection{Relativistic regime}
In the relativistic regime, when $v_0$ becomes comparable to light velocity $c$ or the relativistic diode voltage $U:=q\phi_g/{mc^2}\succsim0.01$, the charge density $\rho$ become 
\begin{equation}
\rho=\gamma \rho_0, \quad \gamma[v]:=\left(1-\frac{v^2}{\hat c^2}\right)^{-1/2}.
\end{equation}
The 1D pressureless Euler-Poisson equations for relativistic electron flow are changed into
\begin{align}
\label{cont2}
\partial_t\rho+\partial_x(\rho v) & =0, \\
\label{relmot}
\gamma^3(\partial_t v+v\partial_x v) &=\partial_x\phi, \\
\label{Poisson2}
\partial_x^2\phi & =\rho. 
\end{align}
Note that the relativistic equations are under the same scaling of Eq.~\eqref{scaling}. Therefore the light speed $\hat c={c T}/{L}$. For computational purpose, let $L/ T=10^7 {\rm m/s}$, so the light speed $\hat c=30$ in this scale.  
The details of the derivation are given in Appendix~\ref{Append1}. In the relativistic regime, the space-charge limit for time-invariant injection is~\cite{Feng2008}
\begin{equation}
 J_{\rm rm}=\frac{c^3}{2d^2} \, \Big[G (U, \gamma_0)\Big]^2,
\end{equation} 
where 
%$G(U, \gamma_0):=\int_0^U{\rm d}u([(u+\gamma_0)^2-1]^{1/2}-(\gamma_0^2-1)^{1/2})^{-1/2}$, 
$G(U, \gamma_0):=\int_0^U \left( \sqrt{(u+\gamma_0)^2-1}-\sqrt{\gamma_0^2-1}\right)^{-1/2} {\rm d}u$, 
$U:=\phi_g/c^2$ and $\gamma_0=\gamma[v_0]$. The same approach will be used to find the SC limited current density for various time profiles as in the classical case~\ref{Classical}.

\section{Results}\label{Results}

In this section, the numerical results are used to pinpoint the space-charge limit when the injection is time-varying. To assess the space-charge effect with respect to the conventional Child-Langmuir limit~\eqref{Jm}, we define two characteristic ratios: the scaled charge
\begin{eqnarray}
\label{qbar}
\bar q:=\frac{\int_0^{\tau_{\rm p}} j_A(t) \, {\rm d}t}{\int_0^{\tau_{\rm p}} J_0(t) \, {\rm d}t} , 
\end{eqnarray}
and the scaled current density 
\begin{eqnarray}
\label{jbar}
\bar j:= \frac{\int_{\tau_1}^{\tau_2}j_A(t) \, {\rm d}t}{J_m(\tau_2-\tau_1)}, 
\end{eqnarray}
in which $j_{\rm A}$ represents the current density at the anode, and $\tau_1$ and $\tau_2$ are the starting and finishing time for  $j_{\rm A}$ being positive. 
In the classical case $\bar q=1$ indicates the critical situation that all injected charge has transmitted through the anode at the finishing time $\tau_{\rm p}$. Due to the leftover charge inside the diode, this should generally be unreachable as confirmed in Fig.~\ref{fig:Fig3}. In the relativistic case (see (a, c, e) of Fig.~\ref{fig:Fig4}),  $\bar q$ becomes larger than unity due to Lorentz length contraction~\cite{SchutzGR}. 
Note $\bar j\geq1$ means that SC limited current density with time-varying pulse for each $j_m(t)$ profile is reached or even transcended.

\subsection{Classical case}
\label{classicalNumerical}
\begin{figure*}[h]

\begin{center}
\includegraphics[width=0.48\textwidth]{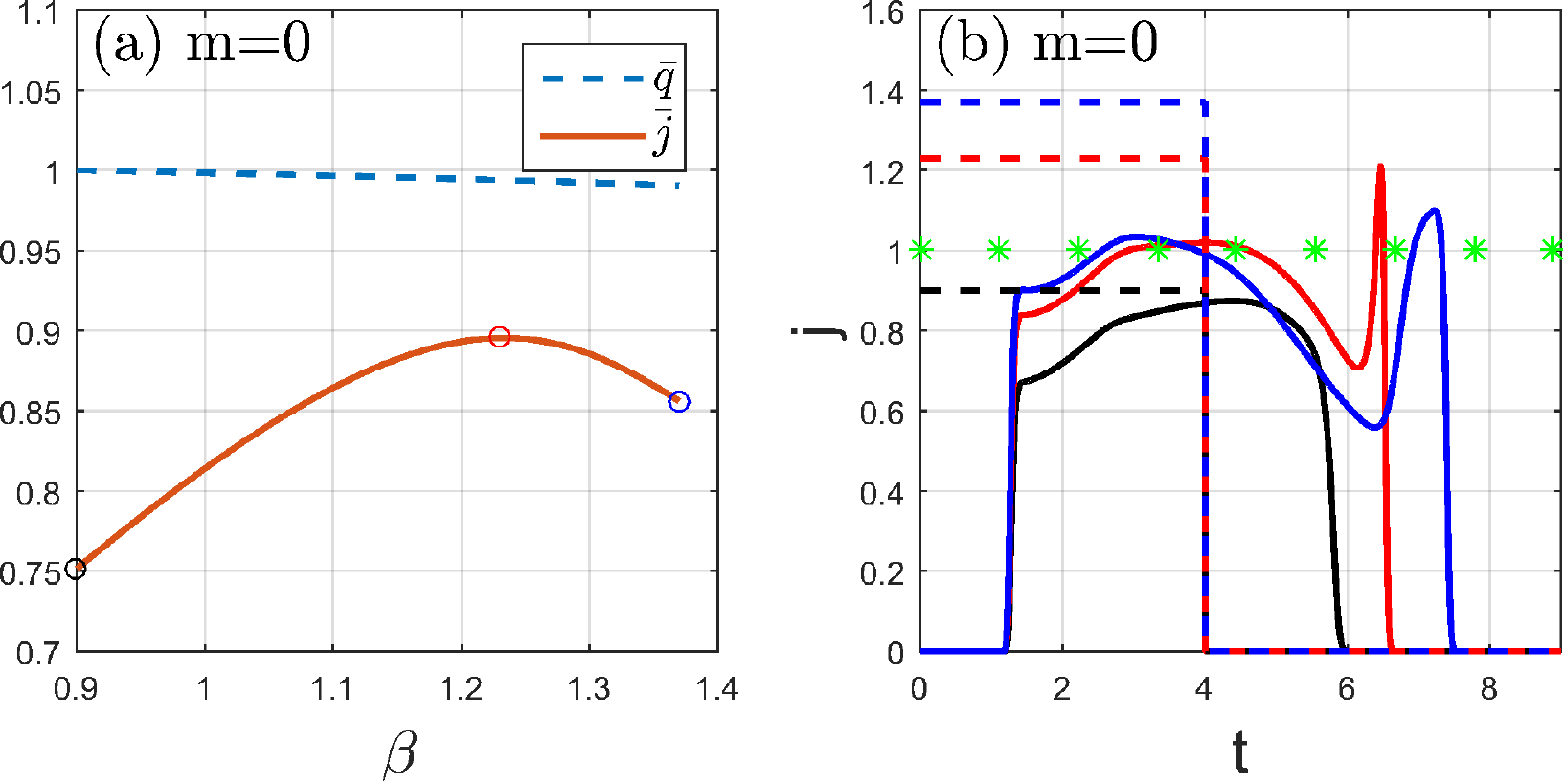}
\includegraphics[width=0.48\textwidth]{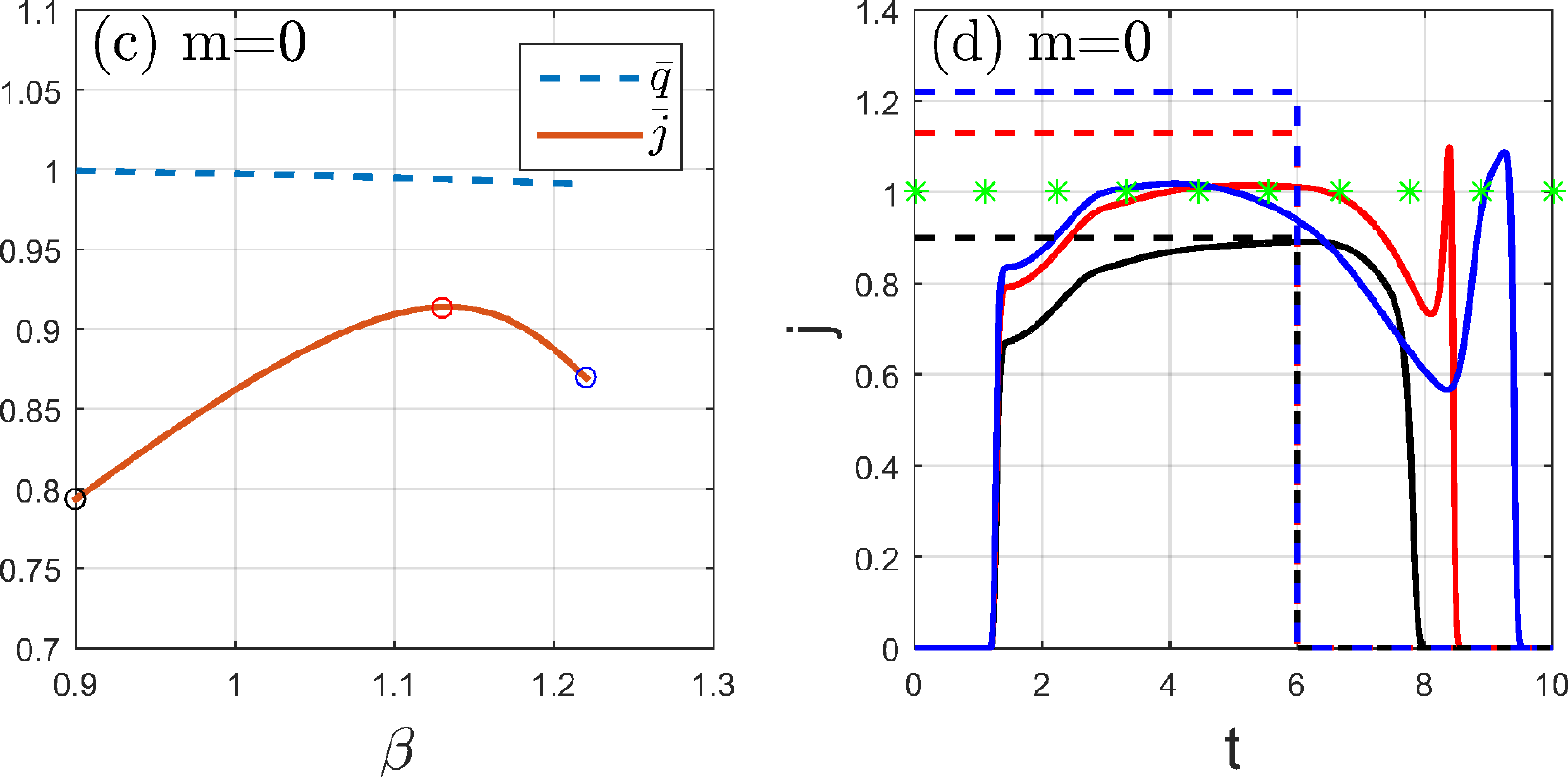}\\

\caption { \label{fig:Fig2} (a, c) Scaled transmitted charge [cf. Eq.~\eqref{qbar}] and scaled current density [cf. Eq.~\eqref{jbar}]. (b, d) Transmittance profiles [solid curves, color marked according to circle marks in (a, c)] at the anode for time-\emph{invariant} injection ones in classical regime. The injection profiles at the cathode are plotted in dashed curves and the SC limit in green asterisks. Parameters: $m=0, v_0=\hat c/60$; (a-b) $\tau_{\rm p}=4$, (c-d) $\tau_{\rm p}=6$. } 

\end{center}
\end{figure*}

We first demonstrate that the output current density $j_{\rm A}(t)$ at the anode is typically time-dependent, 
even when the injection current density is constant over the pulse. 
The numerical results for the case $m=0$ are presented in Fig.~\ref{fig:Fig2}. 
In Fig.~\ref{fig:Fig2} (a), when $\beta_0$ varies from 0.90 to 1.37, the transmitted charge at the anode (dashed line) is almost identical to the emitted charge at cathode. A careful check indicates that the smallest amount is 0.98.  On the other hand in Fig.~\ref{fig:Fig2} (b), as $\beta_m$ increases, the transmitted current density (red dashed) increases to the optimal value 89\% of the SC limited one (red solid line)~Eq.~\eqref{Jm} and then goes down until 86\% (blue dashed line). 
These observations are based on the ratios $\bar{q}$ and $\bar{j}$ defined in Eqs.~\eqref{qbar} and \eqref{jbar}. 
When magnitude $\beta_0$ increases more than 1.37, the whole beam starts to reflect and the space-charge limit is reached. 
We further study three typical cases of $\beta_0$ found from Fig.~\ref{fig:Fig2} (a), 
which are denoted in colored circles.
In Fig.~\ref{fig:Fig2} (b) the emitted currents at the cathode (dash lines) and the transmitted currents at the anode (solid lines) of those three cases are presented in matching colors. 
The pulse length is chosen to be $\tau_{\rm p}=4$ in order to be longer than the time $\tau_1=1.15$ when the anode current becomes positive. 
It is concluded that even when the injection flow is time-invariant, the transmitted flow is time-dependent at the anode. Moreover, when the injection is smaller than space-charge limit (denoted as the green asterisk in Fig.~\ref{fig:Fig2}), the transmitted current is relatively uniform in time (the black curves). 
When the injection is larger, the transmittance becomes damped around the central region and electrons are shaped to high values at the pulse end (see red and blue curves therein). 

When the pulse length is set longer as $\tau_{\rm p}=6$, a larger current density can transmit across the anode, seen in Fig.~\ref{fig:Fig2} (c). 
Similar saturation effect due to space-charge limit is observed for the longer pulse length. 
The transmitted current density is extended in the central part of the pulse, as the anode current density  $j_{\rm A}$ in Fig.~\ref{fig:Fig2} (b, d) show. 
To make a fair comparison, we choose $\tau_{\rm p}=4$ in the rest classical simulations.

\begin{figure*}[h]

\begin{center}
\includegraphics[width=0.48\textwidth]{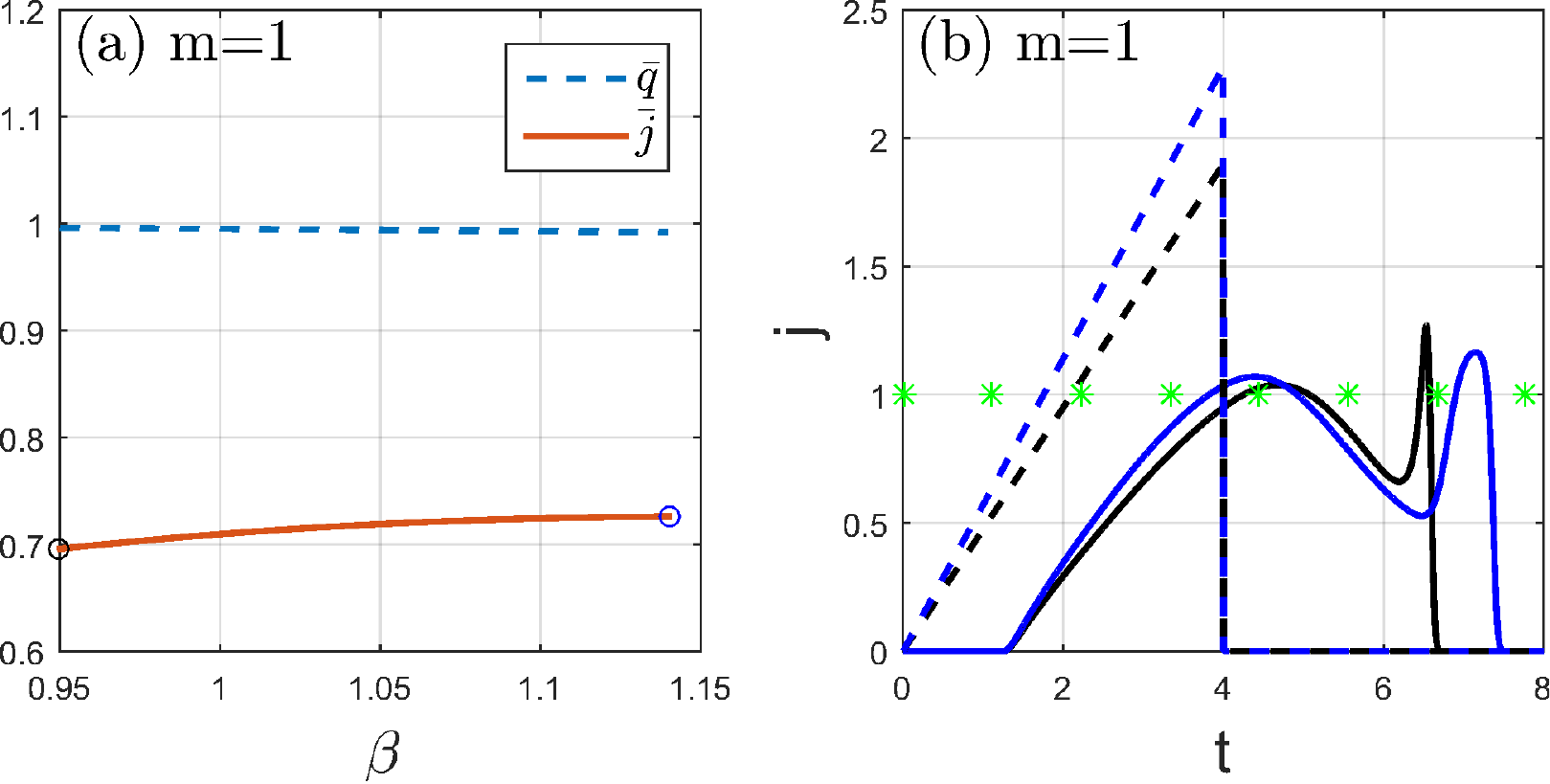}
\includegraphics[width=0.48\textwidth]{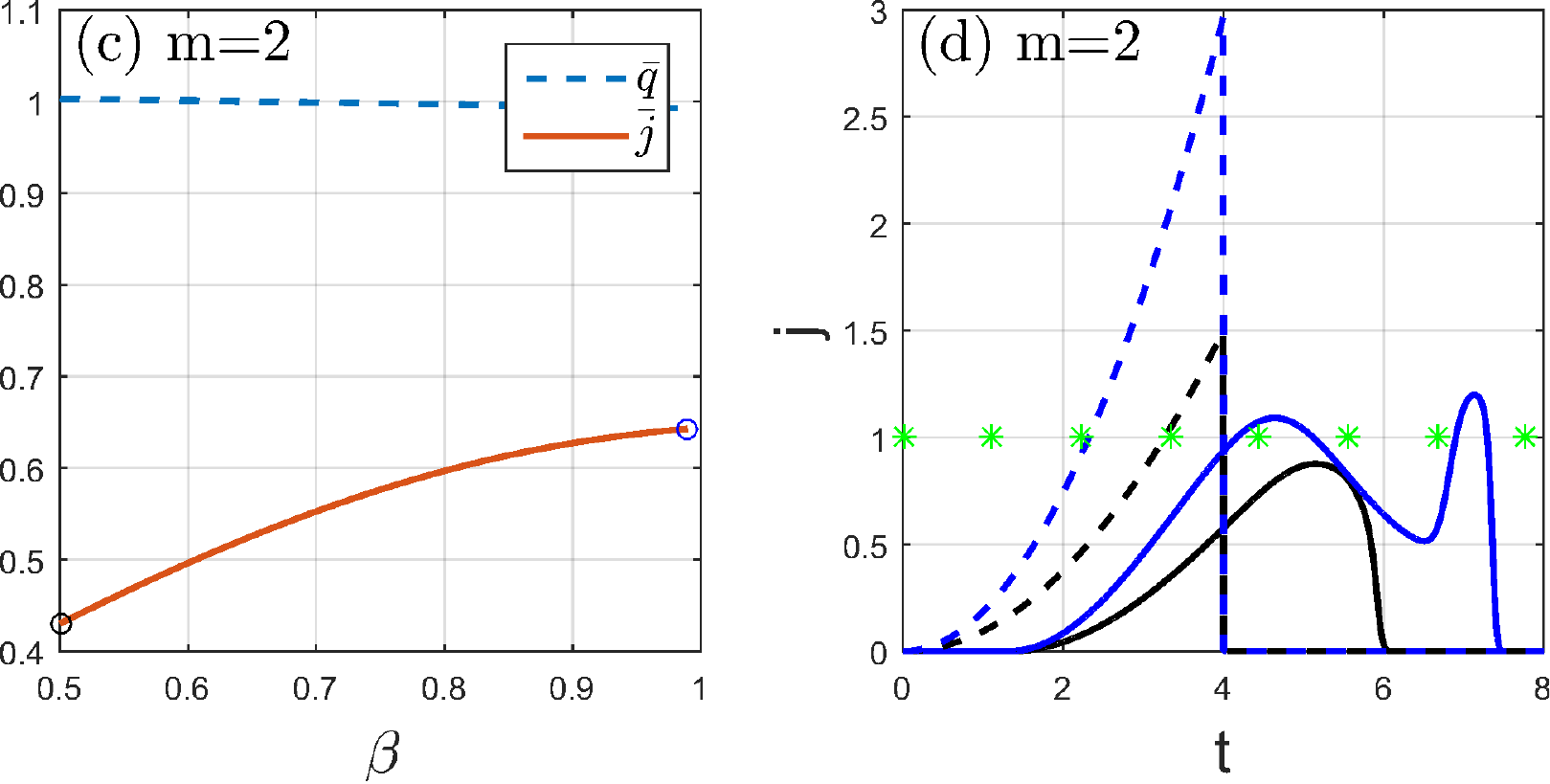}\\
\includegraphics[width=0.48\textwidth]{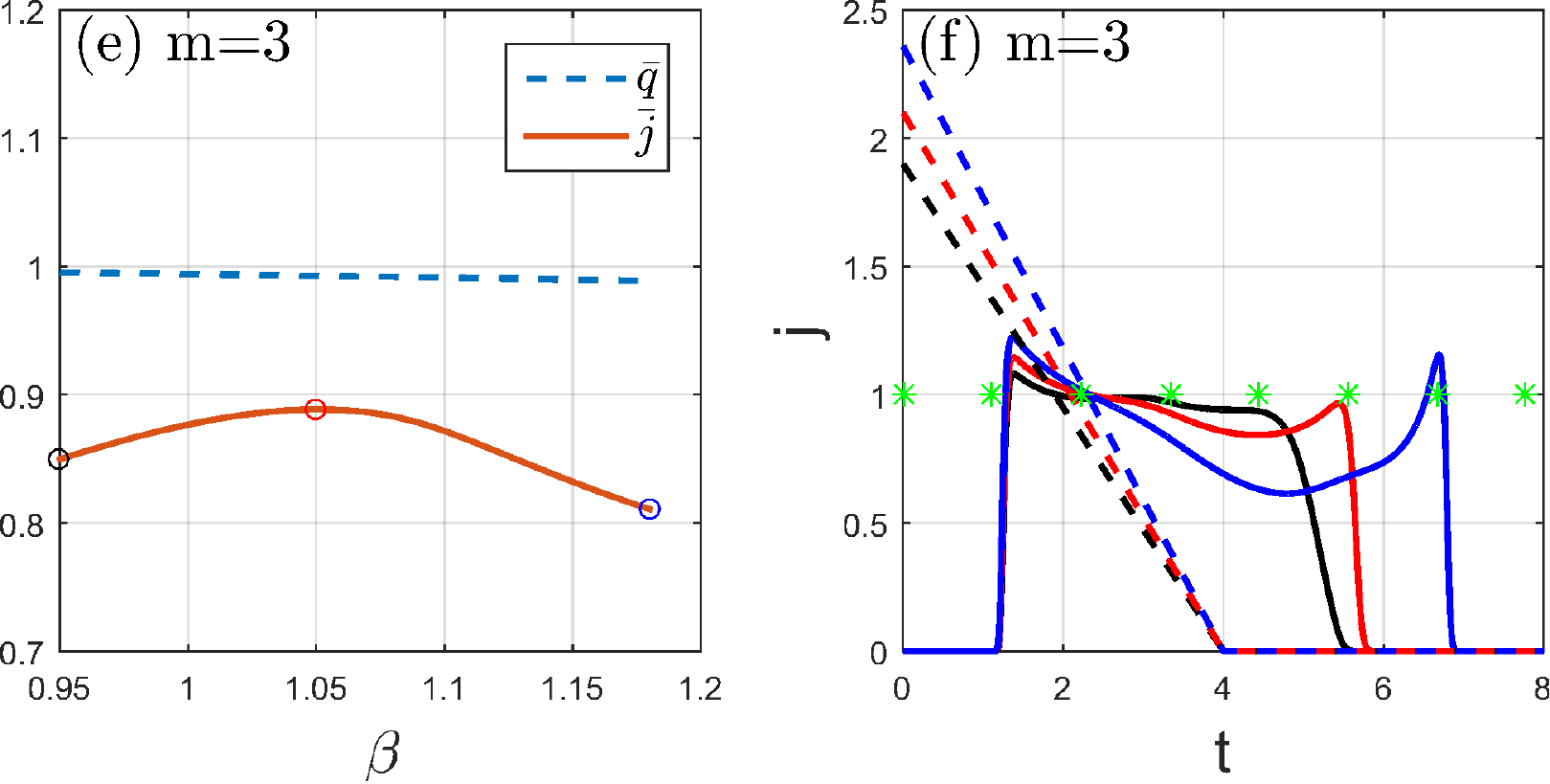}
\includegraphics[width=0.48\textwidth]{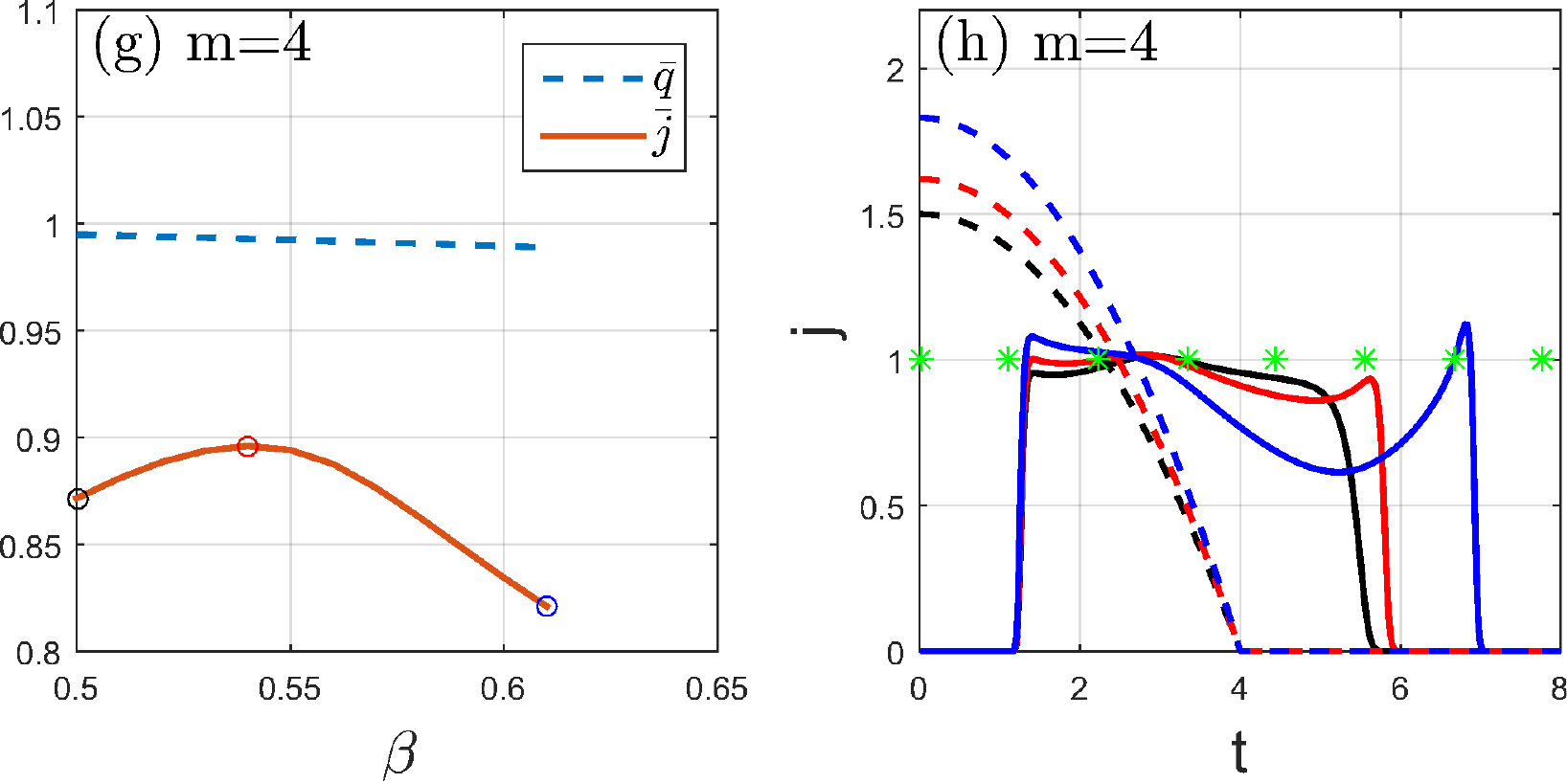}\\
\includegraphics[width=0.48\textwidth]{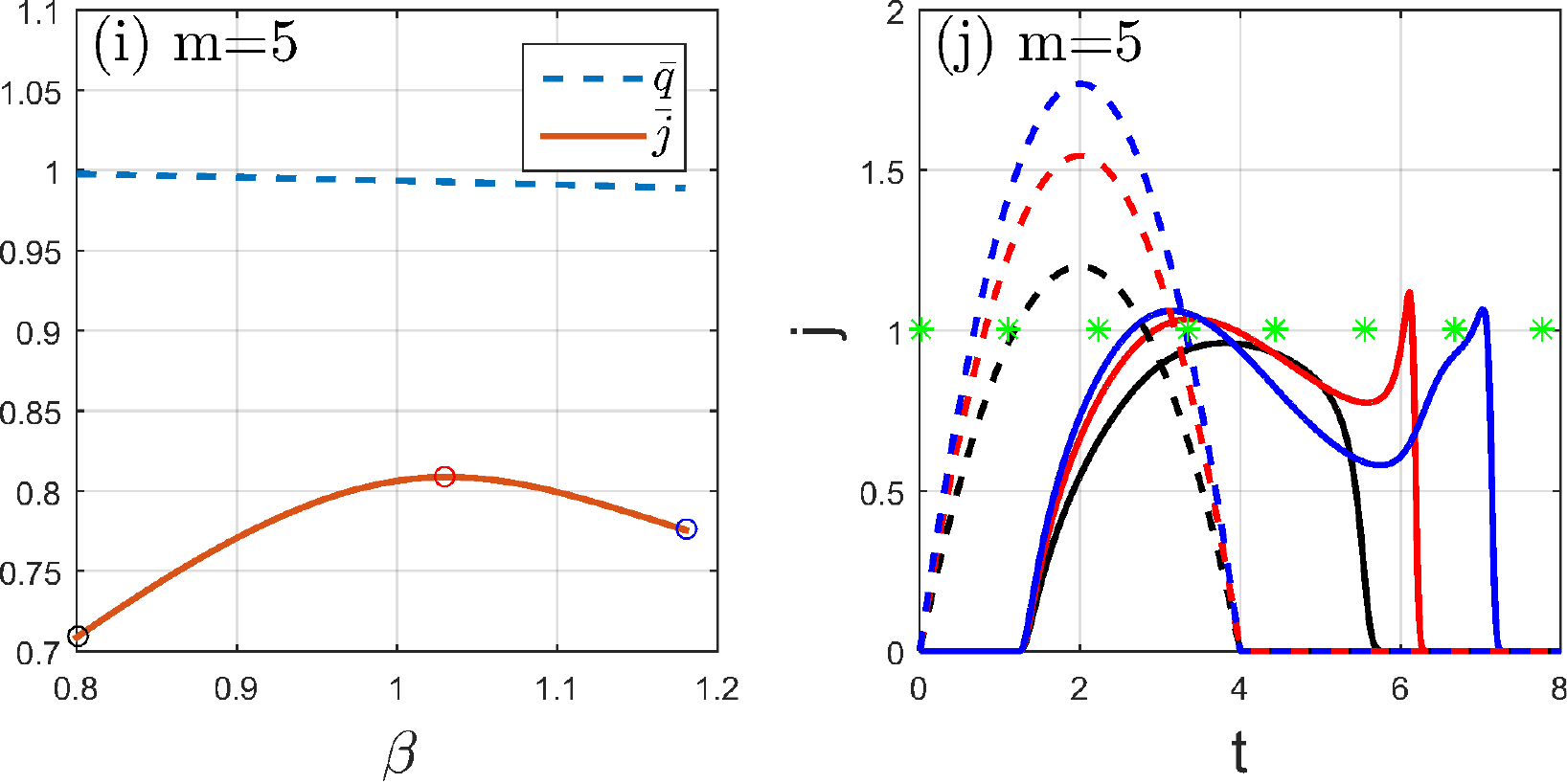}
\includegraphics[width=0.48\textwidth]{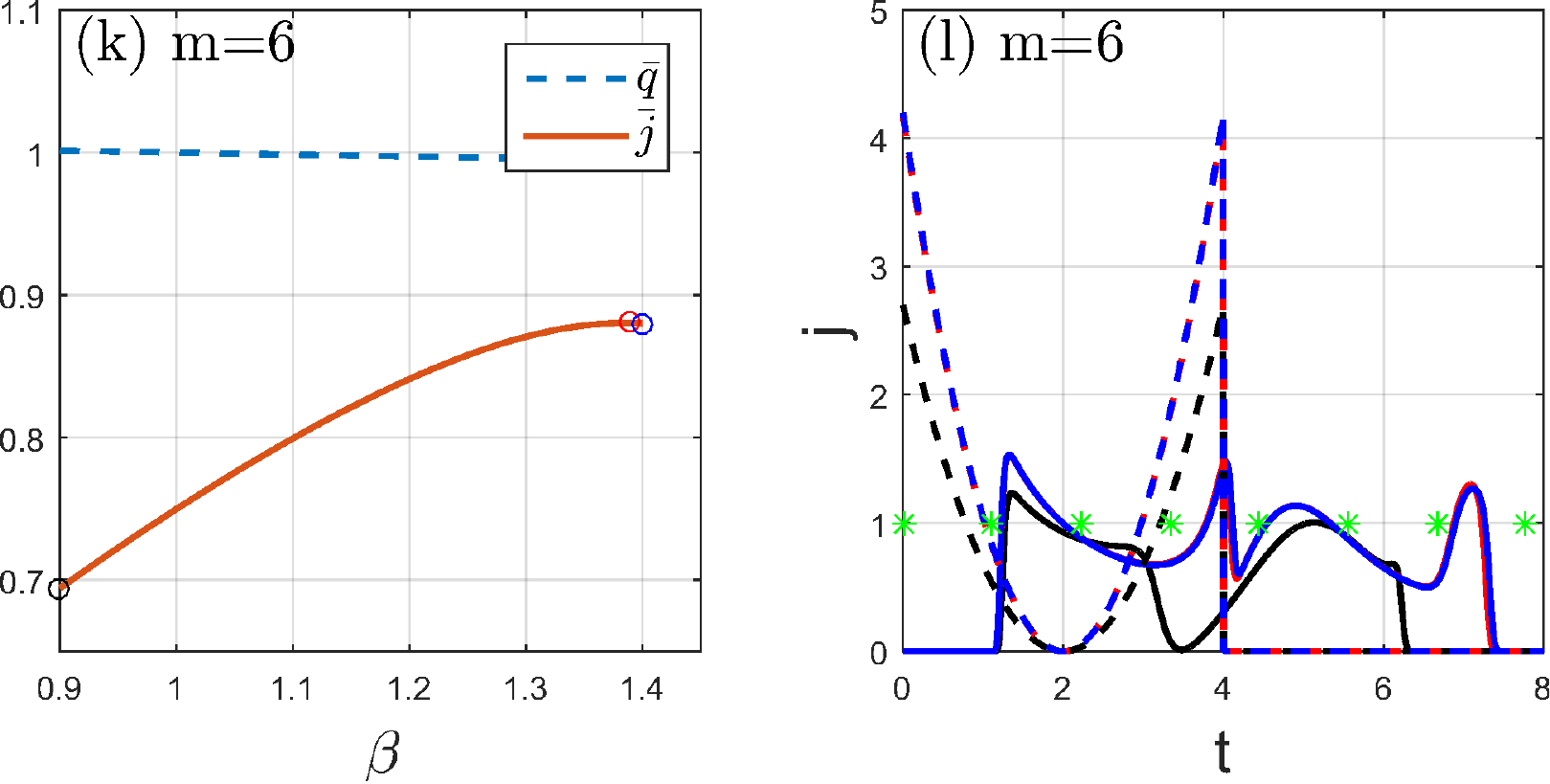}

\caption { \label{fig:Fig3} (a, c, e, g, i, k) Scaled transmitted charge [cf. Eq.~\eqref{qbar}], scaled current density [cf. Eq.~\eqref{jbar}]. (b, d, f, h, j, l) Transmittance profiles at the anode for time-varying injection ones. Other marks are the same as Fig.~\ref{fig:Fig2}. Parameters:  $v_0=\hat c/60, \tau_{\rm p}=4$, (a-b) $m=1$; (c-d) $m=2$; (e-f) $m=3$; 
(g-h) $m=4$; (i-j) $m=5$; (k-l) $m=6$. }

\end{center}
\end{figure*}

Next, we study the time-varying injection of the cases $m= 1 \sim 6$. 
All results for six time-varying injection profiles are plotted in Fig.~\ref{fig:Fig3}(a-l). The output density for time-varying injection is again time-varying, 
%The varying {\color{blue} nature} of injection is also echoed in the output density, 
similar to the constant injection case before. 
For each $m$ there exists an optimal magnitude $\beta_m$ that the transmittance researches the maximum, closest to SC limit in Eq.~\ref{Jm}. 
For example when $m=4$ in Fig.~\ref{fig:Fig3} (g-h), the maximal transmittance reached at $\beta_4=0.54$ (in red circle) is about $90\%$ of the space charge limit. 
If $\beta_4$ is larger than $0.54$, the transmittance decreases.
When $\beta_4>0.62$ (see the blue circle in Fig.~\ref{fig:Fig3} (h)), the electron beam starts to reflect towards the cathode and it is therefore determined that space charge limit has reached. 
The three cases in Fig.~\ref{fig:Fig3} (h) indicates that the transmittance starts to damp in the central region due to the clamping SC effect (the most pronounced for the blue curve). 
It shows that the SC effect dominates the time-dependent electron flow before there would be a larger current density across the diode. 
For all other profiles, similar clamping effects due to space charge repulsion are observed from the results. 
Based on these simulations, we conclude that \emph{time-varying injection seem unlikely to contribute more than the space-charge limit~\eqref{Jm}}. 
This is the main result of this paper.

\subsection{Relativistic case}
\begin{figure*}[h]

\begin{center}
\includegraphics[width=0.6\textwidth]{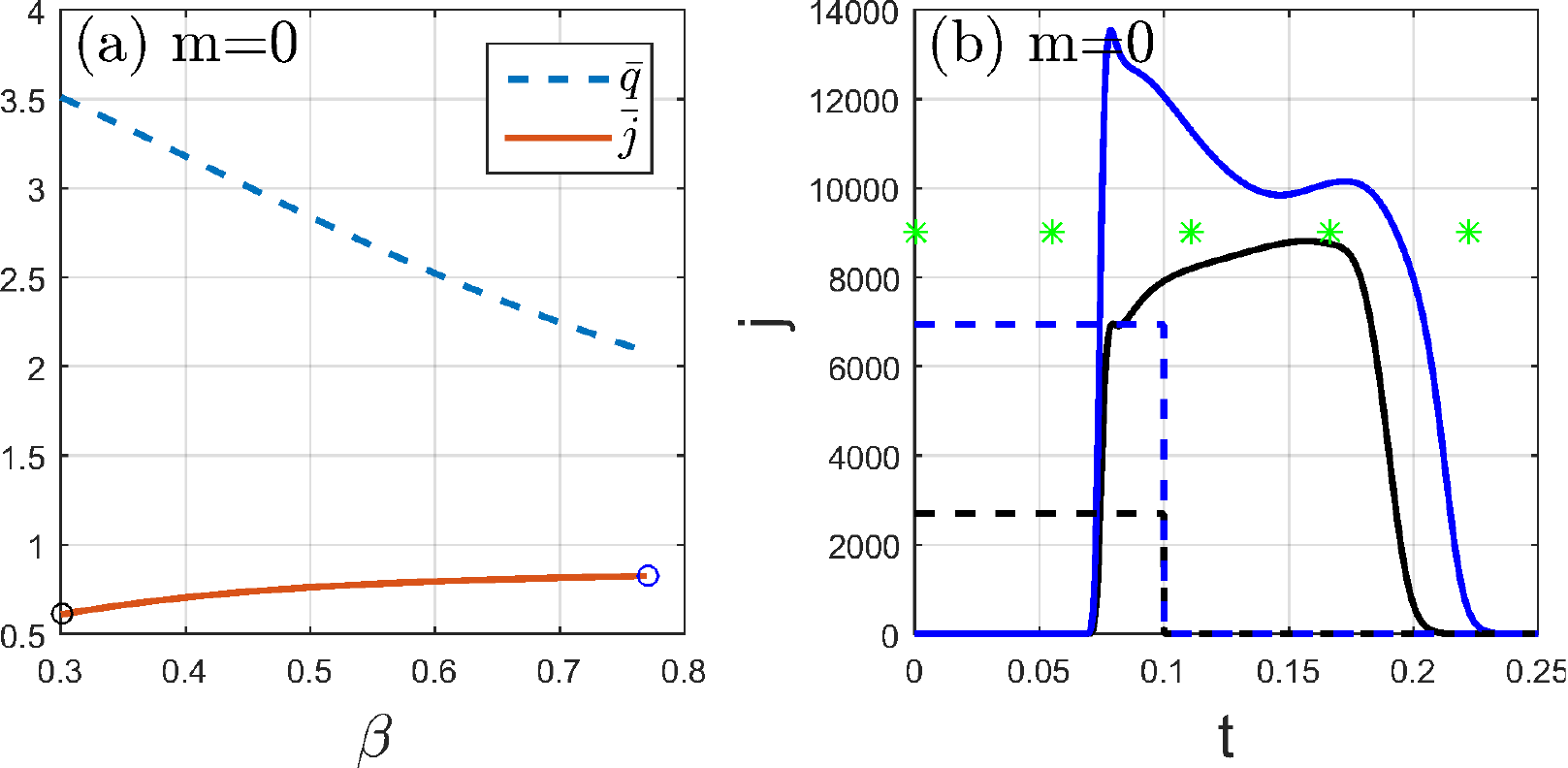}
\includegraphics[width=0.6\textwidth]{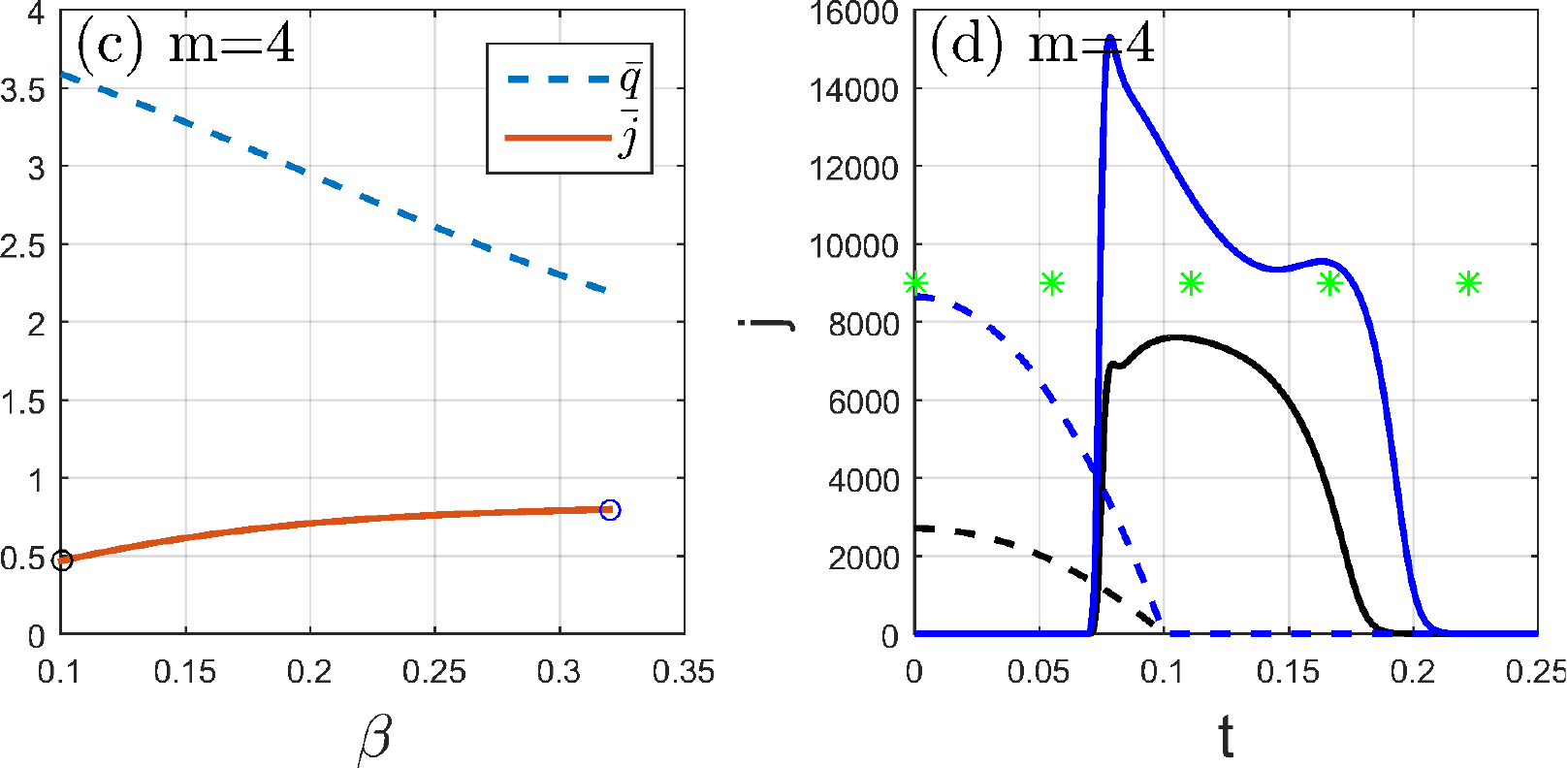}
\includegraphics[width=0.6\textwidth]{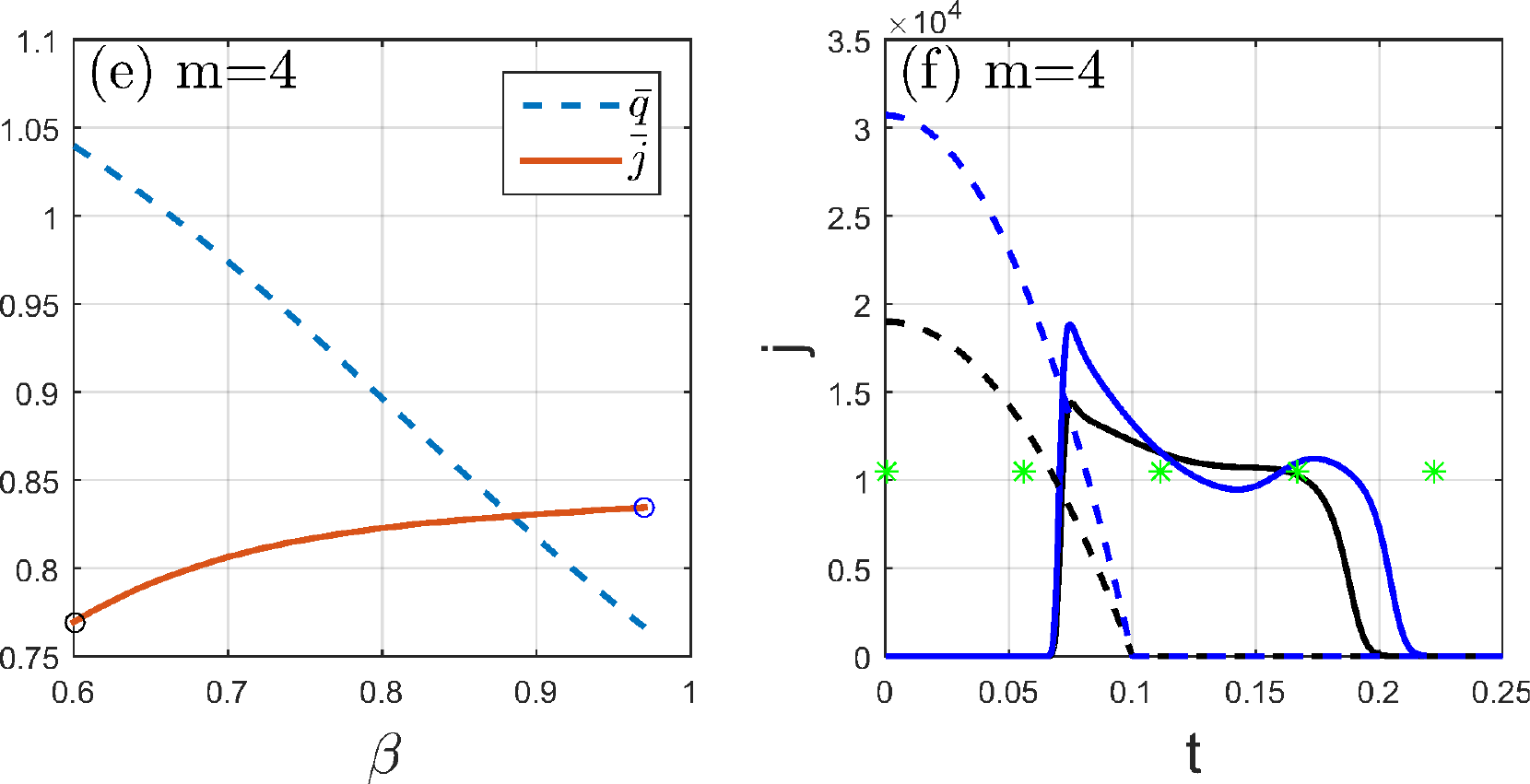}

\caption { \label{fig:Fig4} (a, c, e) Scaled transmitted charge [cf. Eq.~\eqref{qbar}] and scaled current density [cf. Eq.~\eqref{jbar}]. (b, d, f) Transmittance profiles at anode for injection ones in relativistic regime ($m=0, 4$). Other marks are the same as Fig.~\ref{fig:Fig2}. Parameters:  $\tau_{\rm p}=0.1$, $\phi_g=900$.  (a-b) $m=0, v_0=\hat c/60$; (c-d) $m=4, v_0=\hat c/60$;  (e-f) $m=4, v_0=\hat c/10$. }

\end{center}
\end{figure*}

In the relativistic regime, we demonstrate that time-varying injection cannot result in a transmittance higher than its SC limit~\cite{Feng2008}. 
Let the anode voltage $\phi_g=900$, and $U= \phi_g/\hat c^2 = 1$ so that the relativistic effect becomes noticeable. \new{When the injection is time-invariant with $v_0=\hat c/60$, the transmittance goes below the known SC limit (see Fig.~\ref{fig:Fig4} (a-b)). Note again that they are scaled quantities following Ref.~\cite{Caflisch2012}, and all velocities are scaled with respect to $L/T$ so light speed $\hat c=30$. }When the injection becomes time-varying ($m=4$ and $v_0=\hat c/60$ for example), 
the SC limit similarly clamps the transmittance profiles as Fig.~\ref{fig:Fig4} (c-d) shows. 
The SC limit also holds when initial velocity is changed to $v_0=\hat c/10$ 
(see Fig.~\ref{fig:Fig4} (e-f)). 
Fig.~\ref{fig:Fig4} (b, d, f) shows the transmittance profiles become damped at the pulse trail when $\beta_m$ increases. 
Note that $\bar q$ becomes larger than unity because the charge density $\rho$ becomes denser due to the relativistic factor (cf.~$j_A=\rho_A v_A$ in Eq.~\eqref{qbar}), and this increasing effect becomes less significant as the injection magnitude $\beta_m$ increases (see Fig.~\ref{fig:Fig4} (a, c, e)).

\subsection{Time-dependent quantities}

\begin{figure*}[h]

\begin{center}
\includegraphics[width=0.6\textwidth]{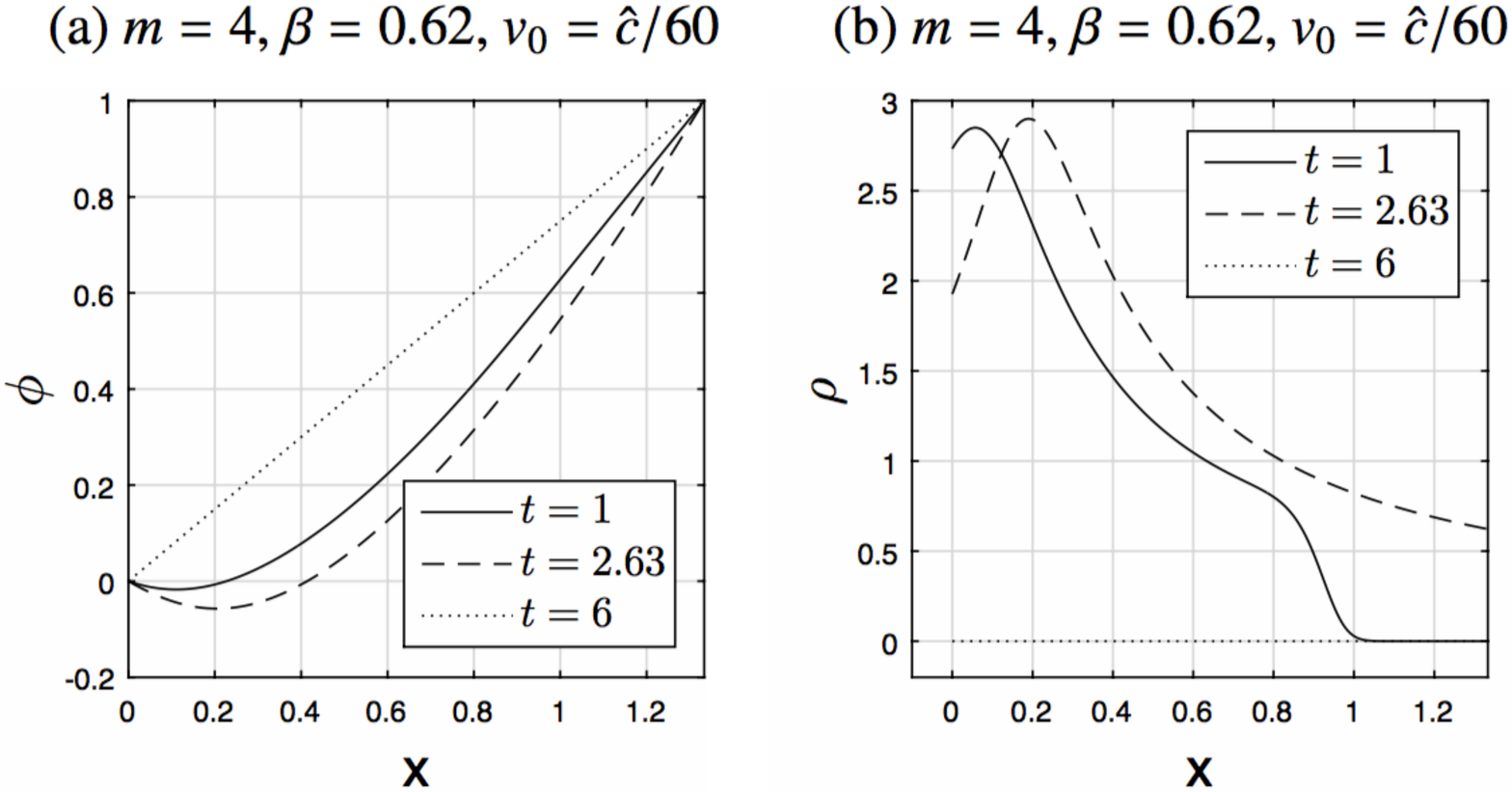}
\includegraphics[width=0.6\textwidth]{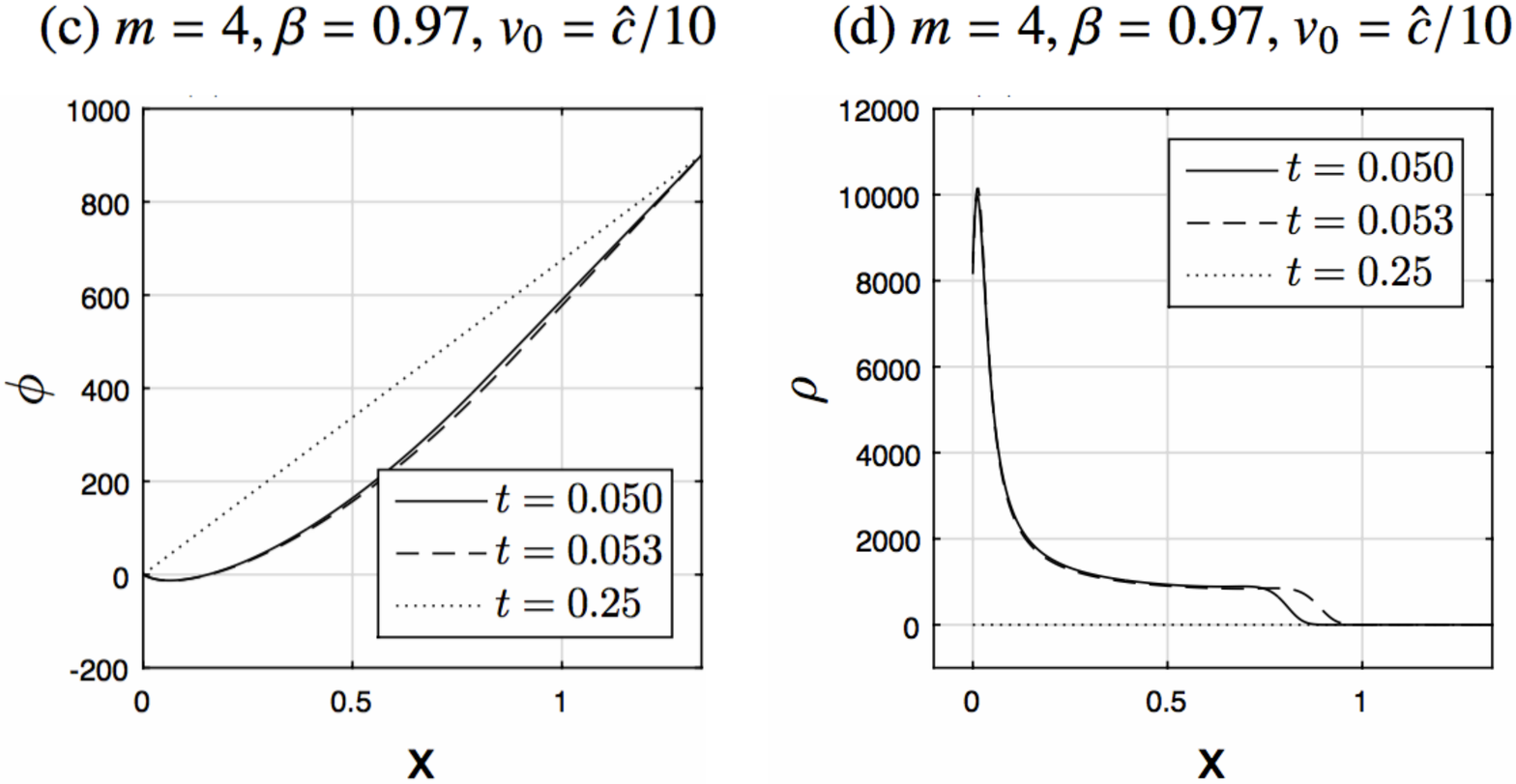}
\caption { \label{fig:Fig5} Electric potential $\phi(x, t)$ and charge density $\rho(x,t)$ for classical and relativistic cases.  Parameters: $m=4$. Classical: (a-b) $\beta_4=0.62$, $v_0=\hat c/60$, $\phi_g=1$; relativistic: (c-d) $\beta_4=0.97$, $v_0=0.1\hat c$, $\phi_g=900$.  }
\end{center}
\end{figure*}

More details can be observed from time-dependent plots of voltage potential $\phi(x, t)$ and charge density $\rho(x, t)$. Fig.~\ref{fig:Fig5} presents $\phi(x, t)$ and $\rho(x, t)$ at three instantaneous time points for a classical case and a relativistic one in Fig.~\ref{fig:Fig5}, when SC limit is reached respectively. We choose $t=2.63$ and $0.053$ to be the time when the injection current density become positive (cf. Figs.~\ref{fig:Fig3}(g-h) and \ref{fig:Fig4}(c-d)) respectively. 
Fig.~\ref{fig:Fig5} shows that as the charge traverses the diode with time, the electric field at the cathode can become positive ($E_{\rm C}=-\partial_x\phi$) and space charge accumulates near cathode to repel more electrons to emit. This is when SC limit dominates the electron flow process and the virtual diode forms. It is also speculated that the zero cathode field $E_{\rm C}=0$ is \emph{not} a critical condition for space charge limit for time-varying injection cases, as previous work~\cite{Jory1969, YL2015} consider so. We speculate that time-dependent space-charge limit may involve more dynamic process of virtual cathode than a static zero cathode field. In our simulation, it is observed that virtual diode oscillation occurs when the scaled $\bar j$ in Eq.~\eqref{jbar} transcends unity. However, we do not include them in Sec.~\ref{Results} because the SC limit is always reached when $\bar j$ remains under unity.

\subsection{Caveat: null emission velocity $v_0=0$}\label{nullv}

%Before we present the results for the general cases when the emission velocity $v_0$ is nonzero, however it is also appropriate to present the special case $v_0=0$. This subsection serves the purpose to address this issue. %\asymp 
%Generally speaking, our simulations should work also for the asymptotic case $0<v_0<<\hat c$. Nevertheless, our algorithm can deal with this issue more accurately. 

\begin{figure*}[h]

\begin{center}
\includegraphics[width=0.8\textwidth]{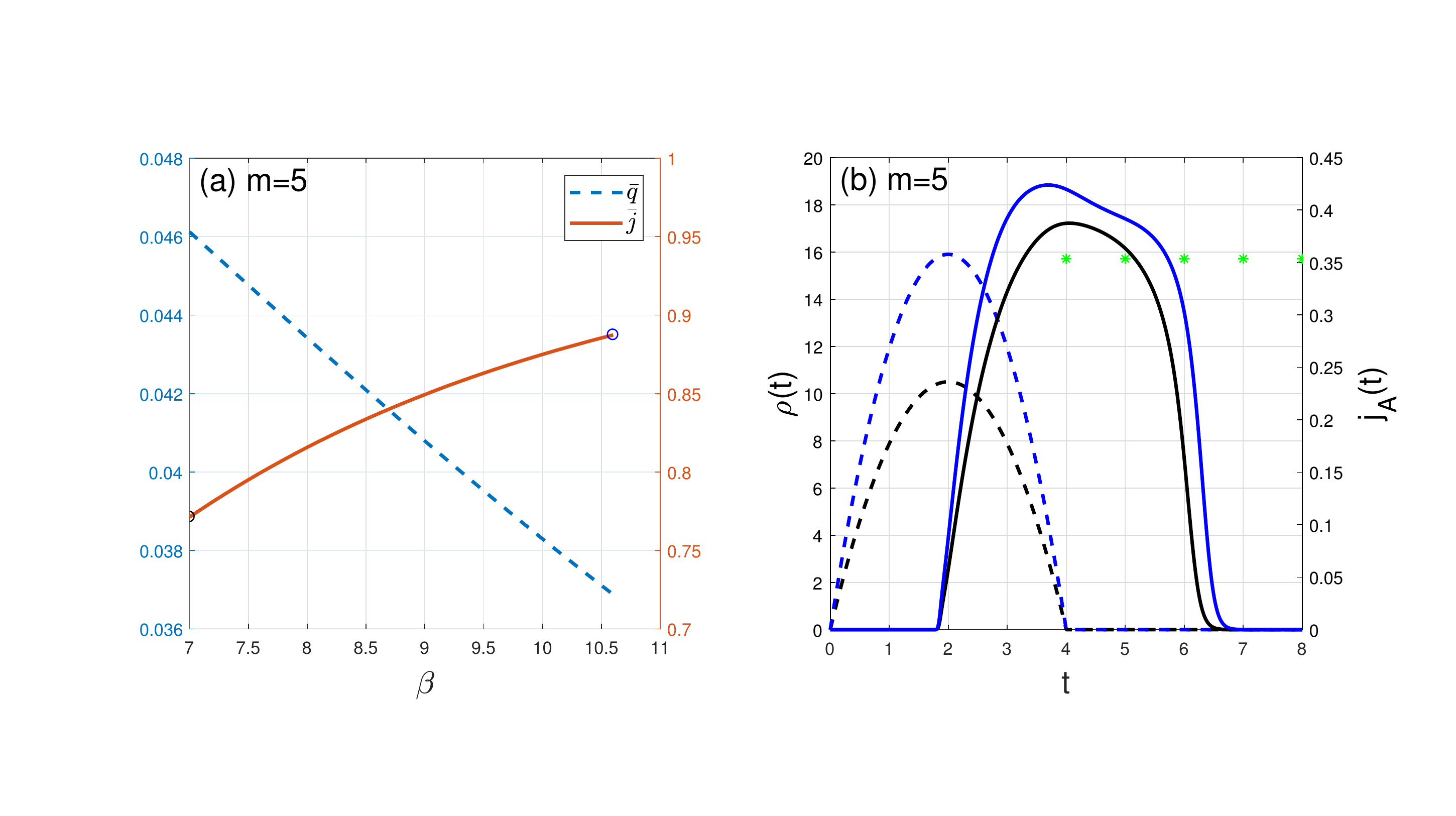}
\caption { \label{fig:Fig6} \new {(a) Scaled transmitted charge [cf. Eq.~\eqref{qbar}] and scaled current density [cf. Eq.~\eqref{jbar}]. (b) Injection charge density $\rho(t)$ (dashed curves) and transmittance profiles at anode $j_A(t)$ (solid curves). Note the curves in (a-b) are plotted in different scales labeled on the left and right sides of the line graph. }Other marks are the same as Fig.~\ref{fig:Fig2}. Parameters:  $m=5, \tau_{\rm p}=4, v_0=0, \phi_g=1$.  [Associated dataset available at http://dx.doi.org/10.5281/zenodo.836163]~\cite{Z2017}} 
\end{center}
\end{figure*}

In the last example, we present the result of the special case when the emission velocity $v_0$ is zero. 
This example is compared with the time varying cases and serves as a validation case since this case is well understood theoretically. The inflow boundary condition Eq.~\eqref{inflow} at $x=0$ becomes 
\begin{align}\label{inflow2}
    \phi=0,  \quad v = v_0,\quad  \rho(t) =\beta_m\sigma_m(t), \qquad 0\le t \le \tau_p, 
\end{align}
where $\beta_m$ is also the dimensionless amplitude and $\sigma_m(t)$ the particular time-dependent profiles assumed for the inflow charge density. \new{Note that we apply a slightly different inflow boundary condition at $x=0$. It is difficult to apply a boundary condition of a finite injection current density with a null emission velocity at $x=0$ in the current model. For such a case, a kinetic model may be necessary to understand its limited current density.} The same outflow boundary condition $\phi=\phi_g$ applies as in the nonzero injection velocity case. The particular profiles $\sigma_m(t)$ are assumed to be polynomial for the same reason as the injection current densities $j_m(t)$ in Eqs.~(8-14).  This is the slightly different inflow boundary condition we apply in this subsection.  Similar to Figs.~\ref{fig:Fig4}, we present a result in Fig.~\ref{fig:Fig6} for $v_0=0, \rho_5(t)=\beta_5\sigma_5(t)$ as the inflow boundary condition, because this represents a reasonable injection current density in pragmatic situations. First in Fig.~\ref{fig:Fig6}(a), as $\beta_5$ increases, the scaled current density increases until 89\% of the SC limit (cf.~blue circle) and most charge remains in the diode spacing at the time of $\tau_{\rm p}$ (cf.~$\bar q<<1$ on the left axis). Second at $\beta_5=10.6$, the transmittance profile is distorted to the most extent (see Fig.~\ref{fig:Fig6}(b)) due to SC effect. Then we show that our simulation also treats the null-velocity case for the time-dependent charge in-flow.

\section{Conclusion}
In summary, this paper addresses the question whether time-varying injection from cathode can push a transmitted current density through anode more than the conventional SC limit. Note that in our paper the injection beam is assumed to be cold and possesses no thermal spread at any instant in time for simplicity~\cite{Lau2001, ZhangP2017}.  From our simulations based on the 1D pressureless Euler-Poisson equations, we speculate that the answer is \emph{negative} under the condition of a fixed voltage difference across the diode. We also extend the classical diode to relativistic regime when initial velocity of electrons is close to the speed of light or when the diode voltage contributes to relativistic acceleration. 

\begin{acknowledgments}

L. Y. was supported by travel grant of Hangzhou Dianzi University [JGB06205-2017031] and UK EPSRC [grant number EP/I034548/1]. He owes the starting idea of time-dependent space-charge effect from L. K. Ang and thanks Lang Jiandong for useful discussion. Q. T. is supported by Eliza Ricketts Postdoctoral Fellowship.

\end{acknowledgments}

%\clearpage

\appendix

\section{Detailed derivation of relativistic extension for 1D electron flow}\label{Append1}

In this section, we start from the full relativistic formulation for the same 1D electron flow problem in Sec.~\ref{Prob}, using covariant four-vector~\cite{Landau1975} and then simplify it to a new differential equations system, Eqs.~\eqref{cont2}, \eqref{relmot} and~\eqref{Poisson2} therein.  There can be many relativistic versions of our problem, for instance see Ref.~\cite{Chan2016}. We choose to formulate in a simple way to be reducible to classical regime when $v\ll c$. 

We use four-vector in SI units (instead of natural units as convention) to represent every vectorial physical quantity in contra-variant components but the final main equations will be scaled  under the same scaling of Eq.~\eqref{scaling}. Hence, the classical current density $j$, vector potential $\mathbf{A}$ and static field $E_x$ are replaced by four current $\mathcal{J}^\mu$, four potential $A^\alpha$, the electromagnetic tensor $F^{\alpha\beta}$ respectively with the metric tensor $g^{\mu\nu}$ and four velocity $U$ being used. The detailed definitions are given as
\begin{eqnarray}
\mathcal{J}^\mu&=&(c\rho, \mathbf{j}),\\
A^\alpha &=& (\frac{\phi}{c}, \mathbf{A}).\\
\partial^\nu:&=&\big(\frac{\partial}{c\partial t}, -\nabla\big), \\
\partial_\nu:&=&\big(\frac{\partial}{c\partial t}, \nabla\big),\\
F^{\alpha\beta}&=&\partial^\alpha A^{\beta}-\partial^\beta A^{\alpha}=\begin{pmatrix}0 & -E_x/c & -E_y/c & -E_z/c \cr
E_x/c & 0 & -B_z & B_y \cr
E_y/c & B_z & 0 & -B_x \cr
E_z/c & -B_y & B_x & 0
\end{pmatrix},\\
g^{\mu\nu}&=&{\rm diag} (1, -1, -1, -1). \\
U^\alpha&=&\gamma(c, \mathbf{v}), \\
U_\alpha&=&\gamma(c, -\mathbf{v}). 
\end{eqnarray}
Therefore, the covariant formulation for the electron flow problem in the three equations~\footnote{Note that the proper time $\tau$ is operated in full differential, instead of partial differential because $\tau$ is frame-independent for a certain particle.  } is written as
\begin{eqnarray}
\label{cont}\partial_\mu\mathcal{J}^\mu&=&0, \\
\label{redyn}
\frac{{\rm d}U^\alpha}{{\rm d}\tau}&=&\frac{q}{m}F^{\alpha\beta}U_{\beta},\\
\partial_\alpha F^{\alpha\beta}&=&\mu_0 \mathcal{J}^{\mu}
\end{eqnarray}
We choose Coulomb gauge so that Poisson equation remains unchanged~\cite{Jackson1999}
\begin{eqnarray}
\label{Poisson}
\nabla^2\phi&=&\frac{\rho(x, t)}{\epsilon_0}, \\
(\nabla^2-\frac{1}{c^2}\partial_t^2)\mathbf{A}&=&-\mu_0 \mathbf{j}_t. 
\end{eqnarray}
Since only 1D situation is considered, $\mathbf{j}=\hat{x}j$ and $\nabla\times\mathbf{j}=0$ means the transverse current vanishes $j_t=0$~\cite{Jackson1999}. It is legitimate to choose 
\begin{equation}
\mathbf{A}=0.
\end{equation}
Then the magnetic field also vanishes, which is consistent with that 1D time-dependent electric field induces no magnetic field from Faraday's law of induction~\cite{Jackson1999}. Eq.~\eqref{redyn} leads to
\begin{equation}
\label{vec2}
\frac{{\rm d}(\gamma v)}{\gamma{\rm d}\tau}=\frac{q}{mc}(-\partial_x\phi)c=\frac{q}{m}F^{10}c=\frac{q}{m}E_x. 
\end{equation}
It is noted that the equation Eq.~\eqref{vec2} differs from $\gamma \partial_t v=q E_x/m$. From Eq.~\eqref{vec2} and $\gamma{\rm d}\tau={\rm d}t$ we have the scaled relativistic dynamical equation [cf.~ Eq.~\eqref{scaling}] as
\begin{eqnarray}
\frac{\rm d}{{\rm d}t}\big[\gamma(v)\cdot v(x, t)\big]=\partial_x\phi, 
\end{eqnarray}
which is expanded as
\begin{eqnarray}
\label{redyn2}
\gamma ^3\frac{\rm d}{{\rm d}t}v(x, t)=\partial_x\phi. 
\end{eqnarray}
The sign appears reverse because of the negative sign of $q$. Eqs.~\eqref{cont}, \eqref{redyn2} and~\eqref{Poisson} can be written as Eqs.~\eqref{cont2}, \eqref{relmot} and~\eqref{Poisson2} $\square$.

%for bibtex, M. added this. 
\bibliographystyle{aipnum4-1}
%\bibliographystyle{apsrev4-1-M}


%\end{thebibliography}%

%\bibliography{bib21}

\begin{thebibliography}{34}%
\makeatletter
\providecommand \@ifxundefined [1]{%
 \@ifx{#1\undefined}
}%
\providecommand \@ifnum [1]{%
 \ifnum #1\expandafter \@firstoftwo
 \else \expandafter \@secondoftwo
 \fi
}%
\providecommand \@ifx [1]{%
 \ifx #1\expandafter \@firstoftwo
 \else \expandafter \@secondoftwo
 \fi
}%
\providecommand \natexlab [1]{#1}%
\providecommand \enquote  [1]{``#1''}%
\providecommand \bibnamefont  [1]{#1}%
\providecommand \bibfnamefont [1]{#1}%
\providecommand \citenamefont [1]{#1}%
\providecommand \href@noop [0]{\@secondoftwo}%
\providecommand \href [0]{\begingroup \@sanitize@url \@href}%
\providecommand \@href[1]{\@@startlink{#1}\@@href}%
\providecommand \@@href[1]{\endgroup#1\@@endlink}%
\providecommand \@sanitize@url [0]{\catcode `\\12\catcode `\$12\catcode
  `\&12\catcode `\#12\catcode `\^12\catcode `\_12\catcode `\%12\relax}%
\providecommand \@@startlink[1]{}%
\providecommand \@@endlink[0]{}%
\providecommand \url  [0]{\begingroup\@sanitize@url \@url }%
\providecommand \@url [1]{\endgroup\@href {#1}{\urlprefix }}%
\providecommand \urlprefix  [0]{URL }%
\providecommand \Eprint [0]{\href }%
\providecommand \doibase [0]{http://dx.doi.org/}%
\providecommand \selectlanguage [0]{\@gobble}%
\providecommand \bibinfo  [0]{\@secondoftwo}%
\providecommand \bibfield  [0]{\@secondoftwo}%
\providecommand \translation [1]{[#1]}%
\providecommand \BibitemOpen [0]{}%
\providecommand \bibitemStop [0]{}%
\providecommand \bibitemNoStop [0]{.\EOS\space}%
\providecommand \EOS [0]{\spacefactor3000\relax}%
\providecommand \BibitemShut  [1]{\csname bibitem#1\endcsname}%
\let\auto@bib@innerbib\@empty
%</preamble>
\bibitem [{\citenamefont {Child}(1911)}]{Child1911}%
  \BibitemOpen
  \bibfield  {author} {\bibinfo {author} {\bibfnamefont {C.~D.}\ \bibnamefont
  {Child}},\ }\href@noop {} {\bibfield  {journal} {\bibinfo  {journal} {Phys.
  Rev.}\ }\textbf {\bibinfo {volume} {32}},\ \bibinfo {pages} {0492} (\bibinfo
  {year} {1911})}\BibitemShut {NoStop}%
\bibitem [{\citenamefont {Langmuir}(1913)}]{Langmuir1913}%
  \BibitemOpen
  \bibfield  {author} {\bibinfo {author} {\bibfnamefont {I.}~\bibnamefont
  {Langmuir}},\ }\href@noop {} {\bibfield  {journal} {\bibinfo  {journal}
  {Phys. Rev.}\ }\textbf {\bibinfo {volume} {2}},\ \bibinfo {pages} {450}
  (\bibinfo {year} {1913})}\BibitemShut {NoStop}%
\bibitem [{\citenamefont {Zhang}\ \emph {et~al.}(2017)\citenamefont {Zhang},
  \citenamefont {Valfells}, \citenamefont {Ang}, \citenamefont {Luginsland},\
  and\ \citenamefont {Lau}}]{ZhangP2017}%
  \BibitemOpen
  \bibfield  {author} {\bibinfo {author} {\bibfnamefont {P.}~\bibnamefont
  {Zhang}}, \bibinfo {author} {\bibfnamefont {A.}~\bibnamefont {Valfells}},
  \bibinfo {author} {\bibfnamefont {L.~K.}\ \bibnamefont {Ang}}, \bibinfo
  {author} {\bibfnamefont {J.~W.}\ \bibnamefont {Luginsland}}, \ and\ \bibinfo
  {author} {\bibfnamefont {Y.~Y.}\ \bibnamefont {Lau}},\ }\href {\doibase
  10.1063/1.4978231} {\bibfield  {journal} {\bibinfo  {journal} {Applied
  Physics Reviews}\ }\textbf {\bibinfo {volume} {4}},\ \bibinfo {pages}
  {011304} (\bibinfo {year} {2017})}\BibitemShut {NoStop}%
\bibitem [{\citenamefont {Jaffé}(1944)}]{Jaffe1944}%
  \BibitemOpen
  \bibfield  {author} {\bibinfo {author} {\bibfnamefont {G.}~\bibnamefont
  {Jaffé}},\ }\href {\doibase 10.1103/PhysRev.65.91} {\bibfield  {journal}
  {\bibinfo  {journal} {Phys. Rev.}\ }\textbf {\bibinfo {volume} {65}},\
  \bibinfo {pages} {91} (\bibinfo {year} {1944})}\BibitemShut {NoStop}%
\bibitem [{\citenamefont {Liu}\ and\ \citenamefont {Dougal}(1995)}]{LiuS1995}%
  \BibitemOpen
  \bibfield  {author} {\bibinfo {author} {\bibfnamefont {S.}~\bibnamefont
  {Liu}}\ and\ \bibinfo {author} {\bibfnamefont {R.~A.}\ \bibnamefont
  {Dougal}},\ }\href {\doibase 10.1063/1.360593} {\bibfield  {journal}
  {\bibinfo  {journal} {Journal of Applied Physics}\ }\textbf {\bibinfo
  {volume} {78}},\ \bibinfo {pages} {5919} (\bibinfo {year}
  {1995})}\BibitemShut {NoStop}%
\bibitem [{\citenamefont {Jory}(1969)}]{Jory1969}%
  \BibitemOpen
  \bibfield  {author} {\bibinfo {author} {\bibfnamefont {H.~R.}\ \bibnamefont
  {Jory}},\ }\href {\doibase 10.1063/1.1657117} {\bibfield  {journal} {\bibinfo
   {journal} {Journal of Applied Physics}\ }\textbf {\bibinfo {volume} {40}},\
  \bibinfo {pages} {3924} (\bibinfo {year} {1969})}\BibitemShut {NoStop}%
\bibitem [{\citenamefont {Howes}(1966)}]{Howes1966b}%
  \BibitemOpen
  \bibfield  {author} {\bibinfo {author} {\bibfnamefont {W.~L.}\ \bibnamefont
  {Howes}},\ }\href {\doibase 10.1063/1.1707862} {\bibfield  {journal}
  {\bibinfo  {journal} {Journal of Applied Physics}\ }\textbf {\bibinfo
  {volume} {37}},\ \bibinfo {pages} {438} (\bibinfo {year} {1966})}\BibitemShut
  {NoStop}%
\bibitem [{\citenamefont {Feng}\ and\ \citenamefont
  {Verboncoeur}(2008)}]{Feng2008}%
  \BibitemOpen
  \bibfield  {author} {\bibinfo {author} {\bibfnamefont {Y.}~\bibnamefont
  {Feng}}\ and\ \bibinfo {author} {\bibfnamefont {J.~P.}\ \bibnamefont
  {Verboncoeur}},\ }\href {\doibase 10.1063/1.3003071} {\bibfield  {journal}
  {\bibinfo  {journal} {Phys. Plasmas}\ }\textbf {\bibinfo {volume} {15}},\
  \bibinfo {pages} {112101} (\bibinfo {year} {2008})}\BibitemShut {NoStop}%
\bibitem [{\citenamefont {Lindsay}, \citenamefont {Li},\ and\ \citenamefont
  {Chen}(2009)}]{Li2009}%
  \BibitemOpen
  \bibfield  {author} {\bibinfo {author} {\bibfnamefont {P.~A.}\ \bibnamefont
  {Lindsay}}, \bibinfo {author} {\bibfnamefont {D.}~\bibnamefont {Li}}, \ and\
  \bibinfo {author} {\bibfnamefont {X.}~\bibnamefont {Chen}},\ }\href {\doibase
  10.1063/1.3183640} {\bibfield  {journal} {\bibinfo  {journal} {Phys.
  Plasmas}\ }\textbf {\bibinfo {volume} {16}},\ \bibinfo {pages} {073303}
  (\bibinfo {year} {2009})}\BibitemShut {NoStop}%
\bibitem [{\citenamefont {Liu}\ \emph {et~al.}(2014)\citenamefont {Liu},
  \citenamefont {Chen}, \citenamefont {Koh},\ and\ \citenamefont
  {Ang}}]{YL2014}%
  \BibitemOpen
  \bibfield  {author} {\bibinfo {author} {\bibfnamefont {Y.~L.}\ \bibnamefont
  {Liu}}, \bibinfo {author} {\bibfnamefont {S.~H.}\ \bibnamefont {Chen}},
  \bibinfo {author} {\bibfnamefont {W.~S.}\ \bibnamefont {Koh}}, \ and\
  \bibinfo {author} {\bibfnamefont {L.~K.}\ \bibnamefont {Ang}},\ }\href
  {\doibase 10.1063/1.4869732} {\bibfield  {journal} {\bibinfo  {journal}
  {Phys. Plasmas}\ }\textbf {\bibinfo {volume} {21}},\ \bibinfo {pages}
  {043101} (\bibinfo {year} {2014})}\BibitemShut {NoStop}%
\bibitem [{\citenamefont {Luginsland}\ \emph {et~al.}(2002)\citenamefont
  {Luginsland}, \citenamefont {Lau}, \citenamefont {Umstattd},\ and\
  \citenamefont {Watrous}}]{Luginsland2002}%
  \BibitemOpen
  \bibfield  {author} {\bibinfo {author} {\bibfnamefont {J.~W.}\ \bibnamefont
  {Luginsland}}, \bibinfo {author} {\bibfnamefont {Y.~Y.}\ \bibnamefont {Lau}},
  \bibinfo {author} {\bibfnamefont {R.~J.}\ \bibnamefont {Umstattd}}, \ and\
  \bibinfo {author} {\bibfnamefont {J.~J.}\ \bibnamefont {Watrous}},\ }\href
  {\doibase 10.1063/1.1459453} {\bibfield  {journal} {\bibinfo  {journal}
  {Phys. Plasmas}\ }\textbf {\bibinfo {volume} {9}},\ \bibinfo {pages} {2371}
  (\bibinfo {year} {2002})}\BibitemShut {NoStop}%
\bibitem [{\citenamefont {Kumar}\ and\ \citenamefont
  {Biswas}(2008)}]{Kumar2008}%
  \BibitemOpen
  \bibfield  {author} {\bibinfo {author} {\bibfnamefont {R.}~\bibnamefont
  {Kumar}}\ and\ \bibinfo {author} {\bibfnamefont {D.}~\bibnamefont {Biswas}},\
  }\href {\doibase 10.1063/1.2836620} {\bibfield  {journal} {\bibinfo
  {journal} {Phys. Plasmas}\ }\textbf {\bibinfo {volume} {15}},\ \bibinfo
  {pages} {023101} (\bibinfo {year} {2008})}\BibitemShut {NoStop}%
\bibitem [{\citenamefont {Caflisch}\ and\ \citenamefont
  {Rosin}(2012)}]{Caflisch2012}%
  \BibitemOpen
  \bibfield  {author} {\bibinfo {author} {\bibfnamefont {R.~E.}\ \bibnamefont
  {Caflisch}}\ and\ \bibinfo {author} {\bibfnamefont {M.~S.}\ \bibnamefont
  {Rosin}},\ }\href {\doibase 10.1103/PhysRevE.85.056408} {\bibfield  {journal}
  {\bibinfo  {journal} {Physical Review E}\ }\textbf {\bibinfo {volume} {85}}
  (\bibinfo {year} {2012}),\ 10.1103/PhysRevE.85.056408}\BibitemShut {NoStop}%
\bibitem [{\citenamefont {Griswold}, \citenamefont {Fisch},\ and\ \citenamefont
  {Wurtele}(2010)}]{Griswold2010}%
  \BibitemOpen
  \bibfield  {author} {\bibinfo {author} {\bibfnamefont {M.~E.}\ \bibnamefont
  {Griswold}}, \bibinfo {author} {\bibfnamefont {N.~J.}\ \bibnamefont {Fisch}},
  \ and\ \bibinfo {author} {\bibfnamefont {J.~S.}\ \bibnamefont {Wurtele}},\
  }\href@noop {} {\bibfield  {journal} {\bibinfo  {journal} {Phys. Plasmas}\
  }\textbf {\bibinfo {volume} {17}},\ \bibinfo {pages} {114503} (\bibinfo
  {year} {2010})}\BibitemShut {NoStop}%
\bibitem [{\citenamefont {Zhu}\ and\ \citenamefont {Ang}(2011)}]{Zhu2011}%
  \BibitemOpen
  \bibfield  {author} {\bibinfo {author} {\bibfnamefont {Y.}~\bibnamefont
  {Zhu}}\ and\ \bibinfo {author} {\bibfnamefont {L.~K.}\ \bibnamefont {Ang}},\
  }\href@noop {} {\bibfield  {journal} {\bibinfo  {journal} {Appl. Phys.
  Lett.}\ }\textbf {\bibinfo {volume} {98}},\ \bibinfo {pages} {051502}
  (\bibinfo {year} {2011})}\BibitemShut {NoStop}%
\bibitem [{\citenamefont {Griswold}, \citenamefont {Fisch},\ and\ \citenamefont
  {Wurtele}(2012)}]{Griswold2012}%
  \BibitemOpen
  \bibfield  {author} {\bibinfo {author} {\bibfnamefont {M.~E.}\ \bibnamefont
  {Griswold}}, \bibinfo {author} {\bibfnamefont {N.~J.}\ \bibnamefont {Fisch}},
  \ and\ \bibinfo {author} {\bibfnamefont {J.~S.}\ \bibnamefont {Wurtele}},\
  }\href {\doibase 10.1063/1.3671961} {\bibfield  {journal} {\bibinfo
  {journal} {Phys. Plasmas}\ }\textbf {\bibinfo {volume} {19}},\ \bibinfo
  {pages} {024502} (\bibinfo {year} {2012})}\BibitemShut {NoStop}%
\bibitem [{\citenamefont {Liu}\ and\ \citenamefont {Ang}(2012)}]{Liu2012b}%
  \BibitemOpen
  \bibfield  {author} {\bibinfo {author} {\bibfnamefont {Y.}~\bibnamefont
  {Liu}}\ and\ \bibinfo {author} {\bibfnamefont {L.~K.}\ \bibnamefont {Ang}},\
  }in\ \href {\doibase 10.1109/PLASMA.2012.6383820} {\emph {\bibinfo
  {booktitle} {The 39th IEEE International Conference on Plasma Science}}},\
  \bibinfo {series and number} {Session 3P: Computational Plasma Physics
  (poster session ~2)}\ (\bibinfo {year} {2012})\ pp.\ \bibinfo {pages}
  {3P--36}\BibitemShut {NoStop}%
\bibitem [{\citenamefont {Pant}\ and\ \citenamefont {Ang}(2013)}]{Pant2013}%
  \BibitemOpen
  \bibfield  {author} {\bibinfo {author} {\bibfnamefont {M.}~\bibnamefont
  {Pant}}\ and\ \bibinfo {author} {\bibfnamefont {L.~K.}\ \bibnamefont {Ang}},\
  }\href {\doibase 10.1103/PhysRevB.88.195434} {\bibfield  {journal} {\bibinfo
  {journal} {Phys. Rev. B}\ }\textbf {\bibinfo {volume} {88}} (\bibinfo {year}
  {2013}),\ 10.1103/PhysRevB.88.195434}\BibitemShut {NoStop}%
\bibitem [{\citenamefont {Liu}\ and\ \citenamefont {Ang}(2014)}]{Liu2014}%
  \BibitemOpen
  \bibfield  {author} {\bibinfo {author} {\bibfnamefont {Y.}~\bibnamefont
  {Liu}}\ and\ \bibinfo {author} {\bibfnamefont {L.~K.}\ \bibnamefont {Ang}},\
  }\href {\doibase 10.1088/0022-3727/47/12/125502} {\bibfield  {journal}
  {\bibinfo  {journal} {Journal of Physics D: Applied Physics}\ }\textbf
  {\bibinfo {volume} {47}},\ \bibinfo {pages} {125502} (\bibinfo {year}
  {2014})}\BibitemShut {NoStop}%
\bibitem [{\citenamefont {Liu}(2014)}]{MyThesis}%
  \BibitemOpen
  \bibfield  {author} {\bibinfo {author} {\bibfnamefont {Y.}~\bibnamefont
  {Liu}},\ }\emph {\bibinfo {title} {Time-Varying Emission of Electrons and Its
  Relevant {EM} Phenomenon}},\ \href {\doibase
  http://hdl.handle.net/10356/60623} {Ph.D. thesis} (\bibinfo {year}
  {2014})\BibitemShut {NoStop}%
\bibitem [{\citenamefont {Griswold}\ and\ \citenamefont
  {Fisch}(2016)}]{Griswold2016}%
  \BibitemOpen
  \bibfield  {author} {\bibinfo {author} {\bibfnamefont {M.~E.}\ \bibnamefont
  {Griswold}}\ and\ \bibinfo {author} {\bibfnamefont {N.~J.}\ \bibnamefont
  {Fisch}},\ }\href {\doibase 10.1063/1.4939607} {\bibfield  {journal}
  {\bibinfo  {journal} {Phys. Plasmas}\ }\textbf {\bibinfo {volume} {23}},\
  \bibinfo {pages} {014502} (\bibinfo {year} {2016})}\BibitemShut {NoStop}%
\bibitem [{\citenamefont {Liu}\ \emph {et~al.}(2015)\citenamefont {Liu},
  \citenamefont {Zhang}, \citenamefont {Chen},\ and\ \citenamefont
  {Ang}}]{YL2015}%
  \BibitemOpen
  \bibfield  {author} {\bibinfo {author} {\bibfnamefont {Y.~L.}\ \bibnamefont
  {Liu}}, \bibinfo {author} {\bibfnamefont {P.}~\bibnamefont {Zhang}}, \bibinfo
  {author} {\bibfnamefont {S.~H.}\ \bibnamefont {Chen}}, \ and\ \bibinfo
  {author} {\bibfnamefont {L.~K.}\ \bibnamefont {Ang}},\ }\href {\doibase
  10.1063/1.4928586} {\bibfield  {journal} {\bibinfo  {journal} {Phys.
  Plasmas}\ }\textbf {\bibinfo {volume} {22}},\ \bibinfo {pages} {084504}
  (\bibinfo {year} {2015})}\BibitemShut {NoStop}%
\bibitem [{\citenamefont {Zhang}\ \emph {et~al.}(2009)\citenamefont {Zhang},
  \citenamefont {Liu}, \citenamefont {Yang}, \citenamefont {Xing},
  \citenamefont {Shao}, \citenamefont {Xiao}, \citenamefont {Zhong},\ and\
  \citenamefont {Lin}}]{ZhangY2009}%
  \BibitemOpen
  \bibfield  {author} {\bibinfo {author} {\bibfnamefont {Y.}~\bibnamefont
  {Zhang}}, \bibinfo {author} {\bibfnamefont {G.}~\bibnamefont {Liu}}, \bibinfo
  {author} {\bibfnamefont {Z.}~\bibnamefont {Yang}}, \bibinfo {author}
  {\bibfnamefont {Q.}~\bibnamefont {Xing}}, \bibinfo {author} {\bibfnamefont
  {H.}~\bibnamefont {Shao}}, \bibinfo {author} {\bibfnamefont {R.}~\bibnamefont
  {Xiao}}, \bibinfo {author} {\bibfnamefont {H.}~\bibnamefont {Zhong}}, \ and\
  \bibinfo {author} {\bibfnamefont {Y.}~\bibnamefont {Lin}},\ }\href {\doibase
  10.1063/1.3124135} {\bibfield  {journal} {\bibinfo  {journal} {Phys.
  Plasmas}\ }\textbf {\bibinfo {volume} {16}},\ \bibinfo {pages} {044511}
  (\bibinfo {year} {2009})}\BibitemShut {NoStop}%
\bibitem [{\citenamefont {Chen}\ \emph {et~al.}(2011)\citenamefont {Chen},
  \citenamefont {Tai}, \citenamefont {Liu}, \citenamefont {Ang},\ and\
  \citenamefont {Koh}}]{Chen2011}%
  \BibitemOpen
  \bibfield  {author} {\bibinfo {author} {\bibfnamefont {S.~H.}\ \bibnamefont
  {Chen}}, \bibinfo {author} {\bibfnamefont {L.~C.}\ \bibnamefont {Tai}},
  \bibinfo {author} {\bibfnamefont {Y.~L.}\ \bibnamefont {Liu}}, \bibinfo
  {author} {\bibfnamefont {L.~K.}\ \bibnamefont {Ang}}, \ and\ \bibinfo
  {author} {\bibfnamefont {W.~S.}\ \bibnamefont {Koh}},\ }\href@noop {}
  {\bibfield  {journal} {\bibinfo  {journal} {Phys. Plasmas}\ }\textbf
  {\bibinfo {volume} {18}},\ \bibinfo {pages} {{023105}} (\bibinfo {year}
  {2011})}\BibitemShut {NoStop}%
\bibitem [{Note1()}]{Note1}%
  \BibitemOpen
  \bibinfo {note} {In 1D case, the current density $j$ are simply related to
  the current $J$: $j=J/A$ where $A$ is the area.}\BibitemShut {Stop}%
\bibitem [{Note2()}]{Note2}%
  \BibitemOpen
  \bibinfo {note} {The increasing profiles (8-10) are normalised so that within
  the emission pulses they give the same account of charge $\DOTSI \intop
  \ilimits@ _0^{\tau _{\protect \rm p}}j_m(t){\protect \rm d}t=\DOTSI \intop
  \ilimits@ _0^{\tau _{\protect \rm p}}1{\protect \rm d}t=\tau _{\protect \rm
  p}$ and the other profiles are chosen to be as smoothly-varying as
  possible.}\BibitemShut {Stop}%
\bibitem [{\citenamefont {Valfells}\ \emph {et~al.}(2002)\citenamefont
  {Valfells}, \citenamefont {Feldman}, \citenamefont {Virgo}, \citenamefont
  {O'Shea},\ and\ \citenamefont {Lau}}]{Valfells2002}%
  \BibitemOpen
  \bibfield  {author} {\bibinfo {author} {\bibfnamefont {{\'A}.}~\bibnamefont
  {Valfells}}, \bibinfo {author} {\bibfnamefont {D.~W.}\ \bibnamefont
  {Feldman}}, \bibinfo {author} {\bibfnamefont {M.}~\bibnamefont {Virgo}},
  \bibinfo {author} {\bibfnamefont {P.~G.}\ \bibnamefont {O'Shea}}, \ and\
  \bibinfo {author} {\bibfnamefont {Y.~Y.}\ \bibnamefont {Lau}},\ }\href@noop
  {} {\bibfield  {journal} {\bibinfo  {journal} {Phys. Plasmas}\ }\textbf
  {\bibinfo {volume} {9}},\ \bibinfo {pages} {2377} (\bibinfo {year}
  {2002})}\BibitemShut {NoStop}%
\bibitem [{\citenamefont {Schutz}(2011)}]{SchutzGR}%
  \BibitemOpen
  \bibfield  {author} {\bibinfo {author} {\bibfnamefont {B.}~\bibnamefont
  {Schutz}},\ }\href@noop {} {\emph {\bibinfo {title} {A First Course in
  General Relativity}}},\ \bibinfo {edition} {2nd}\ ed.\ (\bibinfo  {publisher}
  {Cambridge University Press},\ \bibinfo {address} {Beijing},\ \bibinfo {year}
  {2011})\BibitemShut {NoStop}%
\bibitem [{\citenamefont {Liu}(2017)}]{Z2017}%
  \BibitemOpen
  \bibfield  {author} {\bibinfo {author} {\bibfnamefont {Y.}~\bibnamefont
  {Liu}},\ }\href {\doibase http://dx.doi.org/10.5281/zenodo.836163.}
  {\bibfield  {journal} {\bibinfo  {journal} {Zenodo}\ } (\bibinfo {year}
  {2017}),\ http://dx.doi.org/10.5281/zenodo.836163.}\BibitemShut {Stop}%
\bibitem [{\citenamefont {Lau}(2001)}]{Lau2001}%
  \BibitemOpen
  \bibfield  {author} {\bibinfo {author} {\bibfnamefont {Y.}~\bibnamefont
  {Lau}},\ }\href {\doibase 10.1103/PhysRevLett.87.278301} {\bibfield
  {journal} {\bibinfo  {journal} {Phys. Rev. Lett.}\ }\textbf {\bibinfo
  {volume} {87}} (\bibinfo {year} {2001}),\
  10.1103/PhysRevLett.87.278301}\BibitemShut {NoStop}%
\bibitem [{\citenamefont {Landau}\ and\ \citenamefont
  {Lifshitz}(1975)}]{Landau1975}%
  \BibitemOpen
  \bibfield  {author} {\bibinfo {author} {\bibfnamefont {L.~D.}\ \bibnamefont
  {Landau}}\ and\ \bibinfo {author} {\bibfnamefont {E.~M.}\ \bibnamefont
  {Lifshitz}},\ }\href@noop {} {\emph {\bibinfo {title} {The Classical Theory
  of Fields}}},\ \bibinfo {edition} {fourth revised english edition}\ ed.,\
  \bibinfo {series} {Course of Theoretical Physics}, Vol.~\bibinfo {volume}
  {2}\ (\bibinfo  {publisher} {Elsevier},\ \bibinfo {address} {Butterworth
  Heinemann},\ \bibinfo {year} {1975})\BibitemShut {NoStop}%
\bibitem [{\citenamefont {Chan}, \citenamefont {Wong},\ and\ \citenamefont
  {Yuen}(2016)}]{Chan2016}%
  \BibitemOpen
  \bibfield  {author} {\bibinfo {author} {\bibfnamefont {W.~H.}\ \bibnamefont
  {Chan}}, \bibinfo {author} {\bibfnamefont {S.}~\bibnamefont {Wong}}, \ and\
  \bibinfo {author} {\bibfnamefont {M.}~\bibnamefont {Yuen}},\ }\href {\doibase
  10.1016/j.jmaa.2016.01.031} {\bibfield  {journal} {\bibinfo  {journal}
  {Journal of Mathematical Analysis and Applications}\ }\textbf {\bibinfo
  {volume} {439}},\ \bibinfo {pages} {925} (\bibinfo {year}
  {2016})}\BibitemShut {NoStop}%
\bibitem [{Note3()}]{Note3}%
  \BibitemOpen
  \bibinfo {note} {Note that the proper time $\tau $ is operated in full
  differential, instead of partial differential because $\tau $ is
  frame-independent for a certain particle.}\BibitemShut {Stop}%
\bibitem [{\citenamefont {Jackson}(1999)}]{Jackson1999}%
  \BibitemOpen
  \bibfield  {author} {\bibinfo {author} {\bibfnamefont {J.~D.}\ \bibnamefont
  {Jackson}},\ }\href@noop {} {\emph {\bibinfo {title} {Classical
  electrodynamics}}},\ \bibinfo {edition} {3rd}\ ed.\ (\bibinfo  {publisher}
  {Wiley},\ \bibinfo {address} {New York},\ \bibinfo {year} {1999})\BibitemShut
  {NoStop}%
\end{thebibliography}

\end{document}